\documentclass[a4paper,english,aps,prl,preprint,superscriptaddress,showpacs,showkeys,longbibliography]{revtex4-2}

\usepackage[utf8]{inputenc}
\usepackage[T1]{fontenc}
\usepackage{lmodern}
\input{glyphtounicode}\pdfgentounicode=1

\usepackage{babel}
\usepackage{mathtools}
\usepackage{amsmath}
\usepackage{amssymb}
\usepackage{graphicx}
\usepackage[colorlinks=true, linkcolor=red, citecolor=blue, urlcolor=blue]{hyperref}
\usepackage[usenames,dvipsnames]{xcolor}
\usepackage{bbold}

\usepackage{soul}
\usepackage{footmisc}

\usepackage{cleveref}

\usepackage{siunitx}
\usepackage{physics} 

\newcommand{\sgn}{\mathrm{sgn}}

\newcommand{\pulse}[2]{\left( #1 \right)_{#2}}
\newcommand{\wait}[2][1em]{\overset{#2}{\rule{#1}{0.4pt}}}
\newcommand{\ind}[1]{_{\mathrm{#1}}}

\begin{document}
	
\title{PHIP Sequences and Dipolar Fields}

\author{Martin C. Korzeczek}
\affiliation{Institut für Theoretische Physik \& IQST, Albert-Einstein Allee 11, Universität Ulm, D-89081 Ulm, Germany}
\author{Ilai Schwartz}
\affiliation{NVision Imaging Technologies GmbH, Wolfgang-Paul Stra{\ss}e 2, 89081 Ulm, Germany}
\author{Martin B. Plenio}
\affiliation{Institut für Theoretische Physik \& IQST, Albert-Einstein Allee 11, Universität Ulm, D-89081 Ulm, Germany}
\email{martin.plenio@uni-ulm.de}

\begin{abstract}
Para-hydrogen induced polarization (PHIP) achieves efficient hyperpolarisation of nuclear spins with the transfer of the singlet order of parahydrogen to target molecules through catalytic hydrogenation reactions and subsequent coherent control of the spin dynamics. However, in realistic conditions B0/B1 inhomogeneities lead to significant reduction in the polarization transfer efficiency. Moreover, in high-concentration samples, dipolar fields arising from the magnetisation of the sample can degrade polarisation transfer efficiency significantly. In this work, we present a theoretical framework and a comprehensive analysis of both pulsed and continuous-wave (CW) control sequences designed to mitigate the detrimental effects of dipolar fields, $B_0/B_1$ inhomogeneities, and moderate chemical shifts. By combining tools from average Hamiltonian theory with detailed numerical simulations, we introduce and characterise a wide range of transfer sequences, including dipolar-field adjusted and dipolar-field suppressing protocols. We identify conditions under which dipolar interactions either hinder or, perhaps surprisingly, stabilise polarization transfer, depending on the sequence structure. Our results offer practical guidance for the selection and design of PHIP transfer sequences under realistic experimental constraints and open pathways toward robust hyperpolarisation in concentrated liquid-state NMR samples.
\end{abstract}

\date{\today }
\maketitle

\section{Introduction}
Nuclear Magnetic Resonance (NMR) is a powerful technique for non-destructive sample characterisation 
and medical imaging. Despite its widespread success, NMR sensitivity remains inherently limited as it relies 
on the weak thermal polarisation of nuclear spins, which, even in strong magnetic fields, typically amounts 
to just a few parts per million at room temperature. This limitation can be overcome through nuclear spin 
hyperpolarisation -- a family of techniques that enhance nuclear spin polarisation well beyond thermal 
equilibrium levels \cite{eillsSpinHyperpolarizationModern2023}. 

Among the wide variety of hyperpolarisation methods, dissolution dynamic nuclear polarisation (dDNP) 
\cite{Ardenkjaer-Larsen_increase_2003} and DNP based on optically polarised triplet spin systems, such as 
pentacene in naphthalene or anthrazene \cite{eichhorn2013hyperpolarization,tateishi2014room,eichhorn2022hyperpolarized} 
and the nitrogen-vacancy centre in diamond \cite{london2013detecting,king2015room,blinder202413,Kavtanyuk2024} have been investigated intensively. 
However, their polarisation rates are limited by the weak effective coupling between the polarisation source and the nuclear target 
spins. This is due to the relatively large spatial separation between the two, as well as a one-to-many relationship between the source and target spins. In contrast, para-hydrogen induced polarisation 
(PHIP) \cite{bowersTransformationSymmetrizationOrder1986,bowersParahydrogenSynthesisAllow1987} achieves direct contact between polarisation source and target, offering the potential for achieving strong hyperpolarisation in suitable molecules. PHIP accomplishes this through an irreversible hydrogenation reaction between a target molecule and para-enriched
hydrogen (parahydrogen) gas. This reaction embeds the nuclear singlet order of parahydrogen into the newly formed
product molecules, achieving a one-to-one ratio between polarisation source and the target. Once the reaction is completed, the singlet order of parahydrogen is converted into observable 
magnetisation on a target nucleus using a variety of methods, such as coherence transfer by NMR pulse sequences 
or adiabatic transfer schemes \cite{bowersTransformationSymmetrizationOrder1986,bowersParahydrogenSynthesisAllow1987,nattererParahydrogenInducedPolarization1997,knechtRapidHyperpolarizationPurification2021}. These control sequences are crucial in determining the effectiveness and robustness of 
polarisation transfer, directly influencing the achievable polarisation. Consequently, their design is 
a key aspect of developing hyperpolarisation technologies.

In this work, we examine the case of low-field heteronuclear PHIP, where the constant magnetic field $B_0$ is sufficiently strong to suppress nonenergy-conserving transitions without inducing a significant differential chemical shift between the two parahydrogen spins embedded in the target molecule via hydrogenation. To transfer the purity of the parahydrogen singlet state to the target spin and generate NMR-active polarisation, magnetic control fields, denoted by $B_1(t)$, are applied to the nuclear spins of the molecule.

For practical applications, it is essential that sequences are robust against spatial inhomogeneities in both $B_0$ and $B_1$. A variety of transfer sequences have been developed to meet these requirements. A comparison of some of these sequences is presented in \cite{korzeczekUnifiedPicturePolarization2023}, where it is also shown that a mathematical equivalence allows the same transfer sequences to be applied to both heteronuclear PHIP where the parahydrogen forms the source of polarisation and pulsed Dynamic Nuclear Polarisation (DNP) where electron spins serve as the source of hyperpolarisation. Similarly, both experimental results and theoretical analysis of specific sequences suggest that a comparable equivalence holds for the case of homonuclear PHIP, where $B_1(t)$ is resonantly driving the hydrogen spins and converts singlet order to hydrogen hyperpolarisation while leaving the heteronuclear state unchanged.

In the case of homonuclear PHIP in low to intermediate strength $B_0$ field, it was recently demonstrated \cite{dagysRobustParahydrogeninducedPolarization2024a} that the performance of existing transfer sequences severely deteriorates with increasing sample concentration. Empirically, an upper limit was found to be determined by the product of sample concentration and sample polarisation, beyond which an increase in sample concentration leads to a corresponding decrease in the achievable polarisation. Identifying the dipolar field created by the sample's own magnetisation as the likely cause of this limitation, a new transfer scheme was devised to overcome this challenge.  This scheme combines the well-known dipolar field decoupling technique of Lee-Goldburg continuous-wave driving \cite{goldburgNuclearMagneticResonance1963,leeNuclearMagneticResonanceLineNarrowing1965} with a modulation that effectively implements the Spin-Locking Induced Crossing (SLIC, \cite{PhysRevLett.111.173002}) transfer sequence in the co-rotating frame. The resulting continuous-wave LG-SLIC transfer sequence successfully enables polarisation transfer at higher sample concentrations, with its theoretical limit determined only by the available $B_1$ field amplitude. 

Continuous-wave control sequences, while energy-efficient, are typically quite sensitive to variations in detuning and power, i.e. $B_0$ and $B_1$ errors, as they rely on achieving energetic resonance. In contrast, pulsed sequences, though consuming more power, tend to be more robust to fluctuations in power and detuning, as they achieve the resonance condition through precise timing of pulses. The enhanced robustness promised by pulsed control sequences is the motivation for this work.

In this work, we provide a more detailed theoretical treatment of the interplay between dipolar field induced dynamics and transfer sequences. We derive and present a broad family of approaches, along with corresponding control sequences -- both pulsed and continuous-wave -- that address the challenges posed by dipolar fields, field inhomogeneities and the presence of small chemical shifts.

\section{Heteronuclear PHIP}
    In heteronuclear PHIP, the purity of the singlet state of para-hydrogen is transferred to nearby heteronuclei on the same molecule to create NMR active hyperpolarisation. For concreteness, we assume that nucleus to be a $^{13}C$ spin label inserted into pyruvate but our considerations are not limited to this case.
    For more succinct equations, we use the convention $\hbar=1$ and describe all parameters as angular frequencies. Spin-operators are described as halved Pauli operators such that e.g. $S_{z}\ket{\uparrow}_S=\hbar/2\,\ket{\uparrow}_S$ with $\ket{\uparrow}_S$ the $z$-aligned positive eigenstate of spin S.
    The DC magnetic field $\vec{B}_0 = B_0\vec{e}_z$ is assumed to be low enough to ensure approximate chemical equivalence between the hydrogen spins $I_1$ and $I_2$ (the equivalence-breaking term described by $\Delta_{CS}$ will be defined in Eq. \eqref{eq:H_err}). The $^{13}C$ spin, denoted $S$, induces a magnetic inequivalence between the hydrogens which serves as the enabler of state transfer from the singlet state. With the heteronuclear $J$-coupling given by $J^{HC}_1 S_z I_{1,z}+J^{HC}_2 S_z I_{2,z}$ it is convenient to define $A=J^{HC}_1-J^{HC}_2$ for the polarisation transfer-relevant coupling and $A^{\Sigma}=J^{HC}_1+J^{HC}_2$ for the polarisation transfer-irrelevant coupling contribution. We assume control fields that are near-resonant to the carbon (hydrogen) Larmor frequencies $\omega_{0,S}$ in $H_{ctrl,S}$ ($\omega_{0,I}$ in $H_{ctrl,I}$), respectively. Using the rotating-wave approximation in the frame co-rotating with the control-field frequency $\omega_{ctrl,S}S_z+\omega_{ctrl,I}(I_{1,z}+I_{2,z})$, the Hamiltonian becomes
    \begin{align}
    H &= J\ \vec{I}_1\cdot\vec{I}_2 + A\,S_z (I_{1,z}-I_{2,z}) + H_{\text{ctrl}}(t) + H_{\text{err}} \label{eq:H}\\ \nonumber
      &\qquad + A^\Sigma\,S_z(I_{1,z}+I_{2,z})
    \end{align}
    with the ideal control field Hamiltonian $H_{ctrl}(t)=H_{ctrl,S}(t)+H_{ctrl,I}(t)$ and error terms collected in $H_{err}$. 
    Unless explicitly stated otherwise, we use parameters corresponding to pyruvate of the form \protect{tert-butyl 4-((2-oxopropanoyl-1-13C)oxy)but-2-ynoate-4,4-d}
    \begin{align}\label{eq:pyruvate_params}
    A& =(2\pi)\,0.4\,\mathrm{Hz}= A^{\Sigma}, \\ \nonumber
    J&=(2\pi)\,11.7\,\mathrm{Hz}, \\ \nonumber
    \Omega &=(2\pi)\,500\,\mathrm{Hz},\\ \nonumber
    \Omega_{I}&=(2\pi)\,600\,\mathrm{Hz},
    \end{align}
    with the hydrogens showing a differential chemical shift of $0.45\,$ppm \footnote{Private communication from S. Knecht at NVision Imaging Technologies GmbH}.
    Here, $\Omega$ and $\Omega_{I}$ are the maximum achievable Rabi frequencies as induced by $H_{ctrl}$:
    \begin{align}
    H_{\text{ctrl}}(t) \label{eq:H_ctrl} =&\, H_{\text{ctrl,S}}+H_{\text{ctrl,I}} \\ \nonumber
        =&\, \vec{\Omega}(t)\cdot \vec{S} + \vec{\Omega}_{I}(t)\cdot(\vec{I}_1+\vec{I}_2) \label{eq:H_err}
    \end{align}
    \begin{align}
    H_{\text{err}} =&\Delta_0 S_z \\ \nonumber
    &+ \Delta_1\ H_{\text{ctrl,S}}/\Omega \\ \nonumber
    &+ \Delta_{DF} (3\langle S_z\rangle S_z-\langle\vec{S}\rangle\cdot \vec{S}) \\ \nonumber
    &+ \Delta_{RD} (\langle S_y \rangle S_x - \langle S_x\rangle S_y) \\ \nonumber
    &+ \Delta_{CS} (I_{1,z}-I_{2,z})
    \end{align}
    The vectorial Rabi frequencies $\vec\Omega(t), \vec{\Omega}_{I}(t)$ represent the Rabi amplitude, phase as well as optional purposeful detuning as part of the vectorial representation.
    For the error terms, $\Delta_0$ describes the detuning between the control field and the carbon spins, S, which typically is created by spatial inhomogeneities in $B_0$ over the sample volume. Similarly, $\Delta_1$ describes amplitude errors in the control field described by $H_{\text{ctrl,S}}$. We do not include error terms for the $I$ spin control field $H_\text{ctrl,I}$ as only a few sequences use this field and the ones that do all use CPMG-like schemes which handle detuning- and amplitude-errors well \cite{maudsleyModifiedCarrPurcellMeiboomGillSequence1986, gullionNewCompensatedCarrPurcell1990}. 
    $\Delta_{RD}$ describes radiation damping \cite{vlassenbroekRadiationDampingHigh1995} from a coil resonant to the $S$ spins. $\Delta_{DF}$ describes the concentration-dependent dipolar field strength, see the next section for a detailed derivation. The development and investigation of sequences that are robust to a significant $\Delta_{DF}$ represents a main focus of this work.

    The term of the largest differential chemical shift between the hydrogen nuclei, $\Delta_{CS}$, to which a sequence remains robust, translates into the largest tolerable field strength $B_0$. For example, with the differential chemical shift of $0.45\,$ppm for the molecule under consideration,
    an absolute chemical shift difference of $\Delta_{CS}=(2\pi)\ 0.3\,$Hz corresponds to $B_0\approx 15\,$mT. We will find that $\Delta_{CS}=(2\pi)\ 0.3\,$Hz corresponds to a safe regime given the parameters in Eq. \eqref{eq:pyruvate_params}. For all sequences that only drive the to-be-polarised $S$-spin, tolerable values of $\Delta_{CS}$ are directly related to the transfer speed of the sequence $A_\ast\le A$. Control sequences that also include pulses on the $I$-spins, i.e. the hydrogen nuclei, can lift this limitation to a certain degree.
    
    Note, on the other hand, that off-resonant coupling of control fields $H_{ctrl}(t)$ can impose lower bounds on the admissible fields $B_0$. In this work, we assume a regime with sufficiently strong $B_0$ where these effects can be neglected.

\section{Overview of sequences and robustness}
	\Cref{tab:seq_overview} provides an overview of all sequences discussed and presented in this work.
    The sequences are grouped into categories based on their relationship to dipolar fields with an additional, separate category for dual-channel sequences. Each entry lists the effective transfer rate $A_\ast/A$ along with the largest single-parameter error for which the corresponding sequence still achieves at least $90\%$ of the full transfer. Values are rounded to the first decimal place. Additionally, robustness plots for single-parameter errors are included for each sequence to offer a visual and more detailed representation of their robustness properties. An enlarged version of this figure will be shown in panel (e) of the figure accompanying the detailed discussion of the relevant sequence.

    Unadjusted sequences such as SLIC and PulsePol become susceptible to dipolar fields when $\Delta_{DF}\sim A$
    whereas adjusted sequences maintain robust polarisation transfer up to $\Delta_{DF}\sim J$ while also preserving other robustness properties. While these adjusted sequences offer some degree of robustness to dipolar field effects, amplitude-swept SLIC is an adiabatic sequence for which the presence of a moderate $\Delta_{DF}$ strength actually enhances the robustness of polarisation transfer.
    Effective suppression of dipolar coupling on the time scale of the $J$ coupling used for polarisation transfer can be achieved using the "DF-suppressing (1)" approach. This does not rely on a single, well-defined $\Delta_{DF}$ value and allows for retaining most of the other robustness properties of polarisation sequences. However, the added dipolar field robustness, limited by the condition $\Delta_{DF}\sim J$, comes at the cost of a slightly reduced transfer rate $A_\ast$.

    Under "DF-suppressing (2)", we categorise sequences that suppress dipolar interactions directly on the $\Omega$ timescale and are therefore not limited by $J$.  Sequences that follow the $J$ precession for creating the transfer interaction can achieve very strong $\Delta_{DF}$ suppression. However, similar to SLIC, they are susceptible to detuning errors $\Delta_0$. MREV-PulsePol combines fast dipolar field suppression on the $\Omega$ timescale with the PulsePol mechanism on the slower $J$ timescale to reach both robustness to $\Delta_{DF}$ and $\Delta_0$ that is not limited by $J$. This robustness, however comes at a certain cost of reduced tolerance to $\Delta_{DF}$ for a given available Rabi amplitude $\Omega$.

    Dual channel sequences further extend these possibilities to allow for higher transfer rates $A_\ast$ in combination with strong detuning robustness to errors in $\Delta_0$. PP+XY is "DF-unadjusted" and shows strong properties with respect to all other robustness metrics. M2A-PP+XY and DF-PP+XY fall under the "DF-suppressing (1)" category and accordingly improve on the $\Delta_{DF}$ robustness up to the $J$ scale. Finally, altMREVpol+XY exemplifies a sequence that offers robustness to all error sources under consideration. Its performance depends on the available Rabi amplitudes $\Omega$, $\Omega_I$ and is not limited by $J$. However, under the Rabi amplitudes under consideration its robustness does not surpass that of the other sequences.

    In practice, when deciding on the sequence most suited to a given experiment, both the molecular parameters $A,J$ and the available Rabi amplitudes $\Omega,\Omega_I$ can differ significantly from those used here. Thus, while the values shown in the table can provide some guidance as to which of the discussed sequences best fulfill any given specific robustness and transfer speed requirements, the exact values are specific to our choice of parameters.

	\begin{table}
		\centering 
		\begin{tabular}{rrrrrrrr} 
                & $A_\ast/A$ & $\Delta_{\text{DF}}$ & $\Delta_0$ & $\Delta_1$ & $\Delta_{\text{CS}}$ & $\Delta_{\text{RD}}$ & \raisebox{-0.4\height}{\includegraphics[width=0.13\linewidth]{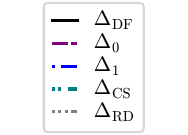}}\\
                \textbf{DF-unadjusted} &  &  &  &  &  & \\
                SLIC & 1.0 & 0.3 & 1.2 & 2.8 & 0.6 & 4.2 & \raisebox{-0.4\height}{\includegraphics[width=0.13\linewidth]{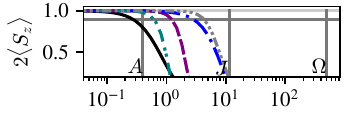}}\\
                PulsePol & 0.7 & 0.2 & 212.7 & 92.7 & 0.5 & 24.5 & \raisebox{-0.4\height}{\includegraphics[width=0.13\linewidth]{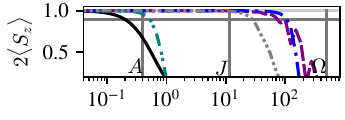}}\\
                \textbf{DF-adjusted} &  &  &  &  &  & \\
                SLIC* & 1.0 & 24.4 & 1.2 & 2.8 & 0.6 & 4.2 & \raisebox{-0.4\height}{\includegraphics[width=0.13\linewidth]{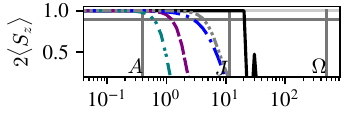}}\\
                PulsePol* & 0.7 & 4.9 & 212.7 & 92.7 & 0.5 & 24.5 & \raisebox{-0.4\height}{\includegraphics[width=0.13\linewidth]{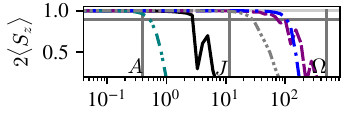}}\\
                \textbf{DF-enabled} &  &  &  &  &  & \\
                amp swept SLIC & 0.3 & 22.9 & 4.6 & 43.7 & 1.0 & 7.2 & \raisebox{-0.4\height}{\includegraphics[width=0.13\linewidth]{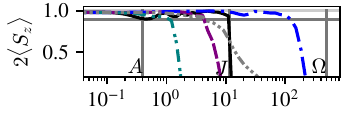}}\\
                \textbf{DF-suppressing (1)} &  &  &  &  &  & \\
                MA-SLIC & 0.8 & 1.6 & 0.1 & 3.4 & 0.5 & 5.3 & \raisebox{-0.4\height}{\includegraphics[width=0.13\linewidth]{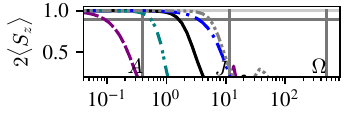}}\\
                MA-PulsePol & 0.6 & 1.2 & 178.0 & 50.8 & 0.5 & 25.1 & \raisebox{-0.4\height}{\includegraphics[width=0.13\linewidth]{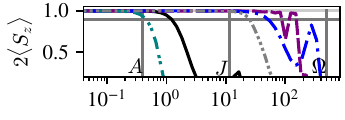}}\\
                M2A-PulsePol & 0.6 & 3.3 & 49.9 & 46.4 & 0.5 & 29.3 & \raisebox{-0.4\height}{\includegraphics[width=0.13\linewidth]{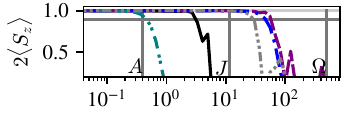}}\\
                DF-PulsePol & 0.5 & 9.0 & 51.9 & 29.5 & 0.4 & 23.6 & \raisebox{-0.4\height}{\includegraphics[width=0.13\linewidth]{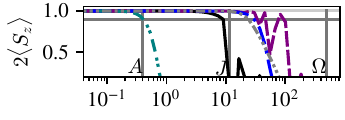}}\\
                \textbf{DF-suppressing (2)} &  &  &  &  &  & \\
                LG-SLIC & 0.6 & 84.7 & 1.7 & 0.9 & 0.4 & 5.1 & \raisebox{-0.4\height}{\includegraphics[width=0.13\linewidth]{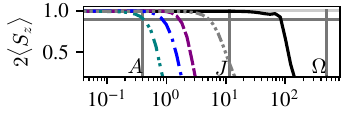}}\\
                BLEWpol & 0.4 & 97.6 & 0.1 & 89.0 & 0.4 & 7.1 & \raisebox{-0.4\height}{\includegraphics[width=0.13\linewidth]{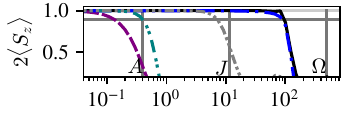}}\\
                MREVpol & 0.5 & 90.8 & 1.7 & 23.8 & 0.4 & 6.0 & \raisebox{-0.4\height}{\includegraphics[width=0.13\linewidth]{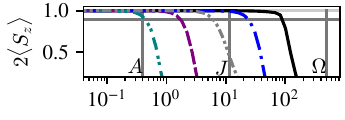}}\\
                MREV-PulsePol & 0.4 & 13.7 & 33.1 & 26.9 & 0.4 & 57.7 & \raisebox{-0.4\height}{\includegraphics[width=0.13\linewidth]{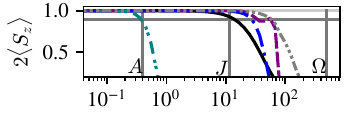}}\\
                \textbf{dual channel} &  &  &  &  &  & \\
                PP+XY & 1.0 & 0.3 & 78.6 & 60.1 & 5.0 & 98.2 & \raisebox{-0.4\height}{\includegraphics[width=0.13\linewidth]{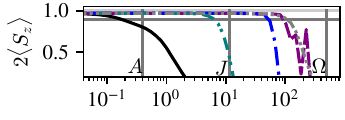}}\\
                M2A-PP+XY & 0.7 & 8.8 & 49.7 & 89.0 & 2.7 & 67.9 & \raisebox{-0.4\height}{\includegraphics[width=0.13\linewidth]{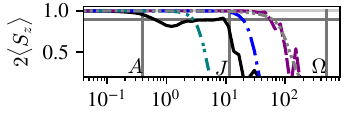}}\\
                DF-PP+XY & 0.6 & 14.2 & 31.5 & 15.4 & 2.4 & 55.2 & \raisebox{-0.4\height}{\includegraphics[width=0.13\linewidth]{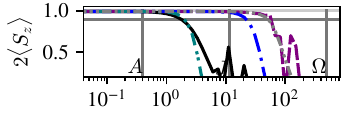}}\\
                altMREVpol & 0.6 & 12.9 & 5.3 & 16.7 & 4.5 & 20.4 & \raisebox{-0.4\height}{\includegraphics[width=0.13\linewidth]{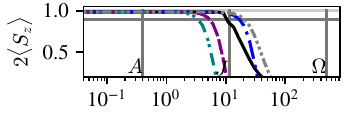}}\\
		\end{tabular}
        \caption{	\footnotesize \label{tab:seq_overview}  Table with an overview over all discussed sequences and their robustness properties. For the error terms, the values given are in units of $(2\pi)\,$Hz such that an entry $0.3$ for $\Delta_{DF}$ corresponds to $\Delta_{DF}=(2\pi)\,0.3$Hz. All entries use $A=(2\pi)\,0.4$Hz, $J=(2\pi)\,11.7\,$Hz, $\Omega=(2\pi)\,500\,$Hz, $\Omega_{I}=(2\pi)\,600\,$Hz which is typical for \protect{tert-butyl 4-((2-oxopropanoyl-1-13C)oxy)but-2-ynoate-4,4-d}. 
        Amplitude swept SLIC uses $\Delta_{DF}=(2\pi)\ 3\,$Hz for evaluating the robustness to other single parameter errors.} 
	\end{table}

    \section{The distant dipolar field}\label{sec:DDF}

    In this section, we derive the nonlinear Hamiltonian description of dipolar fields used in Eq. \eqref{eq:H_err}, starting from the microscopic description of the dipole-dipole interactions between all molecules.
    Apart from notational differences, our description is fully equivalent to established approaches described in the literature, e.g. \cite{devilleNMRMultipleEchoes1979, heIntermolecularMultipleQuantum1993, levittDemagnetizationFieldEffects1996}.

    Our focus is on the spin $S$ contributions to the dipolar fields. In principle, dipolar couplings can also arise from other spin species, such as the hydrogen spins, $I$, or from interactions between different spin species. The former can be neglected when the hydrogen spins remain in the $m=0$ manifold, where the magnetisation vanishes in the mean-field description. The latter contribution is typically small due to the fact that they exhibit merely thermal polarisation which makes their dipolar field contribution negligible.

    In the lab frame, at any moment in time $t$ 
    we can describe our sample as a macroscopic number of molecules with $S$-spins, denoting spin $i$ at position $\vec{r}_i(t)$ with $S_i$. Starting out, we describe the intermolecular dipole-dipole interaction between these spins which for spin $i$ yields the Hamiltonian
    \begin{align}
    H''_{dip, i} = \sum_{j} \dfrac{\mu_0 \gamma_S^2}{4\pi} \dfrac{1}{|\vec{r}_{ij}|^3} \left( \vec{S}_j\cdot\vec{S}_i - 3 \dfrac{(\vec{S}_j\cdot \vec{r}_{ij})(\vec{S}_i\cdot\vec{r}_{ij})}{|\vec{r}_{ij}|^2} \right),
    \end{align}
    where $\vec{r}_{ij} := \vec{r}_j-\vec{r}_i$. In the liquid state motional diffusion will rapidly change the relative orientation and magnitude of $\vec{r}_{ij}=\vec{r}_{ij}(t)$ for nearby molecules. The interplay between translational and rotational diffusion, the dipole-dipole coupling terms and the Larmor precession results in an incoherent net exchange between spins in the sample which gives rise to the intermolecular Nuclear Overhauser Effect (NOE) \cite{solomonRelaxationProcessesSystem1955}. For our purposes this part of the interaction results in a modification of the lifetime of the spins in the sample while further details can be neglected in the description of the coherent dynamics of polarisation sequences. 

    For more distant molecules, diffusion on the time scales of the interaction affects $\vec{r}_{ij}$ negligibly and can be averaged out. For a spin-only description that incorporates the effects of diffusion it is thus natural to describe the spins as a continuous medium $S_i,\vec{r}_i \to S(\vec{r}),c(\vec{r})$ with a concentration $c$. The coherent contributions from dipole-dipole interactions experienced by $S(\vec{r})$ at a 
    position $\vec{r}$ become
    \begin{align}
        H'_{dip}(\vec{r}) &= \int_{\mathcal{V}^*} \mathrm{d}^3r'\ c(\vec{r}\,')\dfrac{\mu_0 \gamma_S^2}{4\pi} \dfrac{1}{|\Delta\vec{r}|^3} \left( \vec{S}(\vec{r}\,')\cdot\vec{S}(\vec{r}) - 3 \dfrac{(\vec{S}(\vec{r}\,')\cdot \Delta\vec{r})(\vec{S}(\vec{r})\cdot\Delta\vec{r})}{|\Delta\vec{r}|^2} \right) \\
        & = \int_{\mathcal{V}^*} \mathrm{d}^3r'\ c(\vec{r}\,')\dfrac{\mu_0 \gamma_S^2}{4\pi} \dfrac{1}{|\Delta\vec{r}|^3} \vec{S}\,^T(\vec{r}\,')\left(  \mathbb{1}- 3 \dfrac{ \Delta\vec{r} \Delta\vec{r}\,^T}{|\Delta\vec{r}|^2} \right)\vec{S}(\vec{r}),\nonumber
    \end{align}
    where we use $\Delta\vec{r} := \vec{r}\,'-\vec{r}$ and denote the geometric volume filled by the sample with $\mathcal{V}$. The asterisk in $\mathcal{V}^*$ indicates that the integration volume leaves out the contributions from nearby spins. The dipolar interaction between pairs of molecules and even their combined effect are weak compared to the Larmor frequency of the spins such that we can discard the energy-nonpreserving contributions. With $\vec{B}_0=B_0 \hat{e}_z$ we reach
    \begin{align}
        H_{dip}(\vec{r}) & = \int_{\mathcal{V}^*} \mathrm{d}^3r'\ c(\vec{r}\,')\dfrac{\mu_0 \gamma_S^2}{4\pi}  \dfrac{3\cos^2\theta -1}{2|\Delta\vec{r}\,|^3} \left(3S_z(\vec{r}\,')S_z(\vec{r}) -\vec{S}(\vec{r}\,')\cdot\vec{S}(\vec{r}) \right), \label{eq:H_dip}
    \end{align}
    where $\cos(\theta) = (\Delta\vec{r}\cdot \hat{e}_z)/|\Delta\vec{r}\,|$. Notably, only the strength—not the form—of the pairwise interaction depends on the positions of the spins. As a result, any sequence that suppresses this type of interaction does so at the level of the microscopic pairwise interaction itself, without relying on the subsequent mean-field approximation.

    If intermolecular coherences are negligible at all times, 
    we can replace the spin operators at position $\vec{r}\,'$ in Eq. \eqref{eq:H_dip} with their expectation values and thus describe the interaction as a classical magnetic field sourced by the spin magnetisation. This field is called demagnetisation field or distant dipolar field in the literature.
    In general, the resulting dynamics can be highly complex or even chaotic \cite{linResurrectionCrushedMagnetization2000}. It can be simplified significantly by assuming spatial homogeneity and a shape of the sample to ensures that $\langle \vec{S}(\vec{r}\,')\rangle$ varies little across the sample \cite{levittDemagnetizationFieldEffects1996}. In the homogeneous case, we set $\langle \vec{S}(\vec{r})\rangle = \langle \vec{S}\rangle$ for all $\vec{r}\in\mathcal{V}$, which yields
    \begin{align}
        H_{DF}(\vec{r}) & = \left(\int_{\mathcal{V}^*} \mathrm{d}^3r'\ c(\vec{r}\,')\dfrac{\mu_0 \gamma_S^2}{4\pi}  \dfrac{3\cos^2\theta-1}{2|\Delta\vec{r}|^3} \right) (3\langle S_z\rangle S_z -\langle \vec{S}\rangle \cdot\vec{S}) \label{eq:H_DF} \\ \nonumber
        &=: \Delta_{DF}(\vec{r}) (3\langle S_z\rangle S_z -\langle \vec{S}\rangle \cdot\vec{S}) \\ \nonumber
        &=: \gamma_S \vec{B}_{DF}\cdot \vec{S}.
    \end{align}
    In the last line, we introduced the sample-induced magnetic field $\vec{B}_{DF}$
    which provides a more natural macroscopic interpretation. In general, the dynamics will not precisely satisfy the assumption of homogeneity used in this final step, but for as long as $H_{DF}(\vec{r})$ does not induce inhomogeneous dynamics across different positions $\vec{r}$, we can disregard the position dependence in  Eq. \eqref{eq:H_DF} and describe the sample using a single representative molecule, $S$.
    
    For this description to remain strictly valid, the polarisation transfer dynamics of $S$ should be insensitive to the precise value of $\Delta_{DF}(\vec{r})$ and the field inhomogeneities ($\Delta_0,\Delta_1$) within the sample volume. A slight generalisation of this condition allows for spatial variation in the magnetisation $|\langle \vec{S}(\vec{r})\rangle|$ while maintaining a homogeneous orientation. This could be formalised by introducing a local correction factor $\chi(\vec{r},t)$ to the homogeneity condition such that $S$ changes its definition to represent the value of $\chi(\vec{r},t)\langle \vec{S}(\vec{r})\rangle$ instead. Although this variation would lead to changes in $\Delta_{DF}(\vec{r})$, it would not alter the general form of Eq. \eqref{eq:H_DF}.

    Assuming a spatially homogeneous concentration $c(\vec{r}\,')=c$, the geometric integral defining the dipolar field strength $\Delta_{DF}(\vec{r})$ averages to zero for a surrounding spherical volume of radius $r_S$. Hence, for any point $\vec{r}$ in the sample only the part of the volume $\mathcal{V}$ that extends beyond the largest solid sphere around $\vec{r}$ gives a net contribution to the dipolar field experienced at that point. Thus the name distant dipolar field. In this work, we use the term dipolar field to emphasise that the relevant properties of dipolar field suppressing polarisation transfer under control sequences 
    are not specific to the liquid state situation that we consider and instead apply to the dipolar couplings of Eq. \eqref{eq:H_dip} in general. The dipolar field adjusted and enabled sequences form \Cref{sec:adjusted} and \Cref{sec:DFenabled}, however, do rely on the mean-field description that leads to the distant dipolar field. Due to the symmetry property and the volume element scaling as $r^2$ in spherical coordinates, the magnitude of $\Delta_{DF}$ depends on the geometrical shape but not on the absolute size as the contribution for a given point is given by $\log R/r_S$ where $r_S$ is the radius of the largest solid sphere that surrounds the regarded point and thus is directly proportional to $R$ itself.
    Additionally it ensures that the detailed definition of the integration volume $\mathcal{V}^*$ compared to the full volume $\mathcal{V}$ does not affect the result.

    An illustrative example is the centre point of a cylinder with its symmetry axis aligned with $\vec{B}_0$ which gives
    \begin{align}
        \label{eq:delta_DF} \Delta_{DF}\langle\vec{S}\rangle /\hbar &= x \gamma_S^2 \mu_0 c \langle\vec{S}\rangle \approx (2\pi)\ 10\,\mathrm{Hz}\ \dfrac{c}{\mathrm{mol/l}} \ \dfrac{\langle\vec{S}\rangle}{\hbar/2}
    \end{align}
    with $x$ a geometric factor which takes the values $x=+1/3$ for the limit of an infinitely long cylinder, and $x=-2/3$ for the limit of an infinitely flat cylinder. The numerical value assumes the long cylinder and the gyromagnetic ratio of ${}^{13}$C. The appendix of \cite{vlassenbroekMacroscopicMicroscopicFields1996} shows that the dipolar field in ellipsoidal geometries is constant, i.e. $\Delta_{DF}(\vec{r})=\Delta_{DF}$ throughout the volume. More detailed treatments of dipolar fields can also be found in e.g. \cite{levittDemagnetizationFieldEffects1996, warrenBoundaryLiquidlikeSolidlike1998, wysinDemagnetizationFields2012}.

    For hydrogen magnetisation from homonuclear PHIP, the roughly $4$ fold gyromagnetic ratio as well as the two hydrogens per molecule lead to a $32$ times enlarged dipolar field strength compared to ${}^{13}$C such that dipolar fields are significantly more limiting in homonuclear PHIP \cite{dagysRobustParahydrogeninducedPolarization2024a}. Technically, both of these dipolar field Hamiltonian terms, as well as mixing terms, are present independent of whether the participating spins are hyperpolarised or not, however a low thermal magnetisation ensures that all of these contributions can be safely neglected when describing the dynamics. For the dual-channel sequences that we present in \cref{sec:dual_channel}, significant hyperpolarisation can accumulate for both the ${}^{13}$C as well as the ${}^1$H spins such that our numerical results use Eq.~\eqref{eq:H_DF} with all of the corresponding terms.

    \section{Methods}
    \subsection{Notation for describing sequences}
        The properties of a control sequence are more naturally defined by its effects on the driven spin as described by $S_z^{rot}(t)$ than by the driving field amplitudes. A sequence, defined via the driving field amplitude $\vec{\Omega}(t)$, is usually described as a sequence of rotations $\pulse{\alpha}{\varphi}$ defined below and waiting times $\wait{\tau}$. We describe a control sequence formally as a product of such segments where in actual time the segments will be applied from left to right. 
        A full sequence, starting at $t=0$, is described by the "core" set of segments, enclosed by square brackets $\left[ \dots \right]^N$, where the core segments are repeated $N$ times, indicated by the superscript. The number of repetitions, N, is typically chosen such that the control sequence ends at approximately $t=2\pi/A_\ast$.  
        An optional set of segments before the start and after the end of the control sequence ensures that all sequences polarise along $+\hat{e}_z$. 
        The duration of the square brackets defines a period $T$ for which $\vec{\Omega}(t+T)=\vec{\Omega}(t)$ in the case of repeating sequences, i.e. whenever $N\neq 1$.
        Pulses are denoted as 
        \begin{align*} 
        \pulse{\alpha}{\varphi} := \pulse{\alpha}{\varphi,\Delta(t)=0}^{\Omega(t)=\Omega} \;\;\mbox{where}\;\; \vec{\Omega}(t) = \begin{pmatrix}
                \Omega(t)\cos \varphi \\
                \Omega(t)\sin \varphi \\
                \Delta(t)
                \end{pmatrix}\;\;\mbox{and}\;\; t_{\text{pulse}}=\dfrac{\alpha}{|\vec{\Omega}(t)|}
        \end{align*}
        which describes a pulse with a total rotation by an angle $\alpha$ and thus a duration of $t_{\text{pulse}}$, and phase $\varphi$. For the continuous-wave schemes like SLIC, the used Rabi amplitude $\Omega(t)$ can be specified explicitly, while for pulsed sequences we typically assume that the maximum available amplitude $\Omega(t)=\Omega$ is applied. For continuous wave schemes such as the Lee-Goldburg drive, an on-purpose detuning $\Delta(t)$ can be specified, while in pulsed sequences usually $\Delta(t)=0$. 
        Here, non-vanishing choices of $\Delta(t)$ implicitly introduce a time-dependence into the control field frequency $\omega_{ctrl,1}$.
        As notation for the pulse phase $\varphi$, which determines the pulse rotation axis, we make use of the shorthand $X=0$, $Y=\pi/2$, $\bar{X}=\pi$ and $\bar{Y}=3\pi/2$.
        To describe complex, non-rectangular waveforms, we write e.g. $\pulse{CW}{\varphi}^{\Omega(t)}$ where the overall rotation angle and pulse duration are given in the surrounding text (cf. \cref{sec:seq_SLIC*}).

        Waiting times in between pulses are denoted as
        \begin{align*}
            \wait[2em]{\tau_{\text{wait}}} \;\;\mbox{where}\;\; \vec{\Omega}(t) = 0 \;\;\mbox{and}\;\; t_{\text{wait}} = \tau_{\text{wait}} - t_{\text{pulses}}.
        \end{align*}
        where a vanishing delay time is sometimes denoted by a $\cdot$. Here $\tau_{\text{wait}}$ denotes the waiting time in the limit of ideal, that is instantaneous, pulses. The actual waiting time $t_{\text{wait}}$ takes the duration of pulses $t_{\text{pulses}}$ into account.
        We provide specifics about which pulses belong to which $\tau_{\text{wait}}$ only
        where we deviate from uniformly distributing all pulse durations across all waiting times such that $t_{\text{wait}}/\tau_{\text{wait}}$ is equal for all waiting times.
        For our notation, the length of the line carries no strict meaning so that $\wait[2em]{\tau}\equiv \wait{\tau}$, but we sometimes draw different lengths to emphasise differences in how many pulses are counted to the corresponding waiting times (cf.~\cref{sec:seq_DF_PulsePol}) or the presence of different waiting times $\tau_{\text{wait}}$ as in \cref{sec:seq_MREVpol}.

    \subsection{Numerically determining robustness properties}
    For a concise evaluation of the robustness of the different sequences for the error contributions in Eq. \eqref{eq:H_err}, we choose to activate only a single error term at a time and calculate the resulting final polarisation $\langle S_z\rangle$ after the sequence is finished. Here, the duration or repetition number $N$ of a sequence is chosen to match as close as possible the theoretical full transfer time $2\pi/A_\ast$. In the panels e) for sequence-specific figures such as \Cref{fig:seq_SLIC}, the resulting fractional transfer for each of the errors over a large range of positive values is shown and \Cref{tab:seq_overview} gives the maximum error strength that still results in a transfer above a threshold of $90\,\%$.
    Note that 
    in some situations, an stopping the sequence earlier can be somewhat preferable in the presence of errors or finite $T_1$ as it may allow for reaching the threshold magnetisation at slightly higher error strengths. We do not include this possible adaptation here in order to keep the discussion simple.
    
    These values usually reflect the properties of a sequence well, however there are several limitations to this approach: First, the point values in \cref{tab:seq_overview} can be misleading for sequences with non-monotonous dependence 
    of the transfer fidelity on the error strength. Here, the more detailed plots can be more informative. Second, the restriction to positive values for the errors 
    assumes that the behaviour of the pulse sequence is similar for positive and negative values, which is often but not always the case. Third, a high robustness to the presence of different errors one at a time does not ensure a similarly high robustness to the simultaneous presence of multiple errors. We have aimed to choose sequences for which the behaviour under the presence of multiple errors at once is unsurprising and where to a good approximation the transfer remains strong as long as every individual error is in the safe regime, but a detailed treatment and proof is beyond the scope of this work.
    
    For sequences with an asterisk in their name, such as SLIC*, the sequence is designed to follow the expected evolution of $\Delta_{DF}$.
    For the dipolar-field enabled sequence, amplitude swept SLIC, the presence of some non-zero $\Delta_{DF}$ is assumed when determining the robustness to other errors, as here the presence of $\Delta_{DF}$ is an important factor that enhances sequence properties.

    \subsection{How to read the sequence-describing figures} \label{sec:seq_describing_figs}
    Every sequence described in this work is accompanied with a corresponding figure such as \Cref{fig:seq_SLIC} for Spin-Lock Induced Crossing (SLIC). 
    
    In each of these, panels a–-c) show characterisations of the sequence: a) is a visual representation of the pulse amplitudes $\vec{\Omega}(t)$ over time, where the hue represents pulse phase, the y-axis represents the driving field amplitude, and optionally a bright or dark tint represents positive or negative detuning ($\Delta(t)$ as in \Cref{fig:seq_MASLIC}). 
    Panels b) shows the amplitudes of the decomposition of $S_z^{rot}(t)$ into the Pauli basis which enters the Average Hamiltonian Theory prediction for the effective Hamiltonian (especially $A_\ast$ and the resonance conditions) and pulse errors $\Delta_0$ and $\Delta_1$. Panel c) shows all unique quadratic contributions which enter the AHT prediction for dipolar field $\Delta_{DF}$ and radiation damping $\Delta_{RD}$ terms. For the visualisation we follow \cite{choiRobustDynamicHamiltonian2020} and choose a color-coded amplitude for each independent component.
    Where space allows, the symbolic pulse definition is given. For better readability, sequences with an asterisk in their name as well as amplitude swept SLIC, the panels use $A=(2\pi)\,1\,$Hz, and many sequences use $\Omega=(2\pi)\,50\,$Hz (MREVpol, BLEWpol and the dual-channel sequences are shown with the full amplitude).

    Panel d) shows the $I$ and $S$ expectation values over time during the sequence in the absence of errors, where the final point in time defines $A_\ast$. The reference frame is the one of \Cref{eq:H}. Panel e) then shows the final magnetisation under a single error type as described in the previous section. 
    The error strengths at which the achieved magnetisation drops below the threshold of $90\,$\% are given in \Cref{tab:seq_overview}. For visual guidance, the threshold as well as the relevant scale-defining parameters $A,J,\Omega$ are also shown.

    For some of the sequences, to demonstrate the importance of specific choices in their definition, we additionally show panels which show plots as in d) and/or e) but for a variant sequence. Alternatively, for  continuous variation, we show lines that give the maximum value of $\Delta_{DF}$ that still achieves the threshold magnetisation of $90\,$\% ($80\,$\%) in dependence of the free parameter.  To indicate the influence of the available Rabi amplitude, we show multiple lines where both $\Omega$ and $\Omega_{I}$ are scaled in proportion. Here, the corresponding legend indicates the respective value of $\Omega$ in $(2\pi)\ $Hz.
    
    \subsection{Limitations to the regarded system}
    Although we use the parameters of pyruvate in the form of \protect{tert-butyl 4-((2-oxopropanoyl-1-13C)oxy)but-2-ynoate-4,4-d} in the numerical simulation, we model the molecule as a 3-spin-system only and neglect decoherence/lifetimes. For this to be approximately accurate, a deuterated molecule has to be used. Here however, dynamical decoupling of the latter has been shown to be important for reaching high absolute polarisation levels \cite{marshallRadioFrequencySweepsMicrotesla2023a}. Although the use of continuous-wave decoupling is to a degree independent of the sequences applied to the ${}^{13}$C and, optionally, the ${}^1$H, in practice this significantly increases the importance of minimising crosstalk of the applied driving fields.
    Thus, in order to allow for strong driving fields $B_1$ while minimising the required $B_0$ field strength that suffices for independently addressing the different spin varieties, in practice an important step for optimising polarisation sequences is the use of shaped pulses such as Gaussian pulses \cite{bauerGaussianPulses1984}. However, especially in the presence of strong dipolar fields, a change of the pulse form might change the corresponding robustness of a sequence as this regime typically requires a large fraction of the overall duration to be filled with the pulses.
    Furthermore, for sequences for which the pulse error robustness (corresponding to $\Delta_0,\Delta_1$) is not limited by system-internal parameters $A,J$ but only by the available Rabi amplitude $\Omega$, composite or shaped pulses can further improve the limits compared to the rectangular pulse shapes we regard here. Again, the interplay with dipolar fields and/or crosstalk provides important boundary conditions.
    A further limitation is that our dipolar field description assumes a uniform magnetisation throughout the sample. For instance, in implementations with very high spatially varying detunings $\Delta_0=\Delta_0(\vec{r})$ or amplitude $\Delta_1=\Delta_1(\vec{r})$, this assumption may no longer hold. Even if the sequence used is robust to variations in $\Delta_{0/1}$ and $\Delta_{DF}$, the uniformity may be compromised if $\Delta_{0/1}/\Omega \ll 1$ does not hold everywhere. In such cases, a more complex description of the dipolar field may be required.

    Finally, while our description does not include finite lifetime effects, in practice, finite lifetimes of the spin order will favour the faster sequences, which is the reason why we include $A_\ast/A$ as an important property of each sequence.

    \section{Effective dynamics and error robustness using Average Hamiltonian Theory}
        
    Following \cite{korzeczekUnifiedPicturePolarization2023}, we describe all polarisation sequences as generating magnetisation along the positive z-axis, $+\hat{e}_z$. By applying appropriate initial and final pulses, this representation can be formally applied to any sequence, allowing us to identify more clearly structural similarities among the different polarisation sequences. 

    To extract the effective dynamics of Eq. \eqref{eq:H} under our control field $H_{ctrl}$ we define a co-rotating frame with the unitary transformation $U^{rot}(t)$ via
    \begin{align} \label{eq:U_rot}
        U^{rot}(0) &= \mathbb{1} \\ \nonumber
        \frac{dU^{rot}}{dt}(t) &= -i (H_{ctrl}(t) +J \vec{I}_1\cdot \vec{I}_2) U^{rot}(t) 
    \end{align}
    and we restrict the state space under consideration to the non-magnetised hydrogen states
    \begin{align} \label{eq:I_pseudospin}
        \ket{T_0} &:= \dfrac{\ket{\uparrow}_{I1}\ket{\downarrow}_{I2} +\ket{\downarrow}_{I1}\ket{\uparrow}_{I2}}{\sqrt{2}}\, , \\ \nonumber
         \ket{S_0} &:= \dfrac{\ket{\uparrow}_{I1}\ket{\downarrow}_{I2} -\ket{\downarrow}_{I1}\ket{\uparrow}_{I2}}{\sqrt{2}}\, , \\ \nonumber
         I_z &:= \dfrac{1}{2}(\ket{T_0}\bra{T_0}-\ket{S_0}\bra{S_0})\, , \\ \nonumber
         I_x &:= \dfrac{1}{2}(\ket{S_0}\bra{T_0}+\ket{S_0}\bra{T_0})\, ,
    \end{align}
    where we follow \cite{korzeczekUnifiedPicturePolarization2023} in our definition of the pseudospin $I$. The co-rotating frame Hamiltonian restricted to the hydrogen levels spanned by $I$ is
     \begin{align} \label{eq:H_rot}
        H^{rot}(t) &= A S_z^{rot}(t) (I_x \cos Jt - I_y\sin Jt )+ H_{\mathrm{err}}^{rot}(t)
    \end{align}
    where $S_z^{rot}(t) = U^{rot}(t) S_z (U^{rot}(t))^{\dagger}$. 
    For the restriction of the dynamics to the non-magnetised hydrogen states to be accurate, it is sufficient for the driving fields to be applied to the heteronuclear $S$ spin only, which is typically an excellent approximation owing to the large detuning of the hydrogen nuclei from ${}^{13}$C nuclei in typical $B_0$-field regimes.
    In \cref{sec:dual_channel} we discuss sequences for which this condition is not fulfilled and pulses are applied also to the $I$ spins, but for now focus our discussion on the simpler case.

    \cite{choiRobustDynamicHamiltonian2020} provides a systematic treatment of the properties of a sequence, via the time evolution it induces on $S_z^{rot}(t)$, along with the application of Average Hamiltonian Theory.
    Their discussion includes pulse errors $\Delta_0,\Delta_1$ and dipolar couplings equivalent to our $\Delta_{DF}$, as well as the geometric requirements for $S_z^{rot}(t)$ to suppress these terms. However, this discussion focuses entirely on sensing and decoupling sequences, not polarisation sequences, which changes the main requirement on a sequence. \cite{korzeczekUnifiedPicturePolarization2023} discusses several existing polarisation sequences from this perspective, while the discussion is restricted to pulse errors $\Delta_0,\Delta_1$. 

    Using Average Hamiltonian Theory,
    the dynamics from $H^{rot}(t)$ can be approximated by an effective Hamiltonian given by
    \begin{align}\label{eq:H_eff_AHT}
    H_{eff}^{rot} &\approx \dfrac{1}{T}\int_0^T\mathrm{d}t\ H^{rot}(t),
    \end{align}
    where we typically assume that the sequence is $T$-periodic, i.e. $H^{rot}(t+T)=H^{rot}(t)$ and $U^{rot}(T)=\mathbb{1}$ to ensure that $H_{eff}^{rot}$ is fully independent of time. Corrections to this effective Hamiltonian \cite{brinkmannIntroductionAverageHamiltonian2016} are of the order of $(H^{rot})^2T/2$ such that for $AT\ll 1$, a sequence which cancels all contributions of $H_\mathrm{err}$ to $H_{eff}^{rot}$ is certified  to decrease the influence of error terms by a factor of order $||H_{\mathrm{err}}||T/2$, where the maximal eigenvalue of the error Hamiltonian competes with the suppression time scale $T$.

    The goal of a polarisation sequence then is to achieve an effective time evolution governed by the Hamiltonian
    \begin{align}\label{eq:H_eff_transfer}
        H_{eff}^{rot} &= A_\ast \dfrac{S_{\tilde{x}} I_x - S_{\tilde{y}} I_y}{2}
    \end{align}
    where the effective coupling strength $A_\ast\le A$ should be as large as possible to reduce the impact of finite polarisation lifetimes and the minus sign is due to the initial state being the negative eigenvalue to the $I_z$ operator. $\tilde{x}$ and $\tilde{y}$ describe two directions with the condition of forming a right-handed coordinate system with $\hat{e}_{\tilde{x}} \times \hat{e}_{\tilde{y}} = \hat{e}_z$ in order to induce polarisation transfer which accumulates along the $+z$ axis.
    
    Deviating from the above, some polarisation sequences use $H_\mathrm{ctrl}(t+T)=H_\mathrm{ctrl}(t)$, but have the added effect of producing an effective non-zero net rotation $\alpha$ around the $z$-axis after a single pulse block
    \begin{align} \label{eq:Urot_precession}
    U^{rot}(T) &= \exp(-i\left[  J T\,I_z + \alpha S_z \right])\, , \\ \nonumber
    U^{rot}(t+T) &= U^{rot}(t) U^{rot}(T).
    \end{align}
    In this case, applying Average Hamiltonian Theory over periods of length $T$ as in Eq. \eqref{eq:H_eff_AHT} results in a piecewise constant Hamiltonian that is rotated by $U^{rot}(T)$ in each iteration. As long as the $H_{eff}^{rot}$ for such sequences contains at least one of the terms of Eq. \eqref{eq:H_eff_transfer}, and if the sequence fulfils $JT=\alpha$ with $\alpha \neq 0,\pi$, this precession will induce the same net interaction as Eq. \eqref{eq:H_eff_transfer}, albeit emerging on a time scale slower than $T$.
    
    One way of treating this scenario within the framework of Average Hamiltonian Theory is to introduce a virtual magnetic bias field equivalent to $U^{rot}(T)$ into the Hamiltonian \cite{nielsenSinglespinVectorAnalysis2019}. This allows sequences to remain effective without requiring $T$ to match full periods of $(2\pi)\,J$, as long as $U^{rot}(T)$ is correctly chosen.
    
    Due to this consideration, we always use $T$ as describing the period of $H_{ctrl}$ and the time scale relevant for suppressing error terms in $H_\mathrm{err}$ and provide the value of $\alpha$ for a given sequence where it is non-zero.
    
    Based on this, we can give quantitative estimates for the acceptable error amplitudes for a sequence that does or does not suppress an error term according to AHT: To this end, we first note that the effective interaction $H_{eff}^{rot}$ from Eq. \eqref{eq:H_eff_AHT} has strength $A_\ast\le A = (2\pi)\ 0.4\,$Hz. For an error term $H_{err}$ that is not suppressed by a sequence, we can expect the resulting contribution $H_{eff,err}^{rot}$ to be of similar magnitude as $H_{err}$ and must expect the possibility of detrimental effects for error strengths which challenge the condition 
    \begin{align}
        ||H_{err}||\ll A_\ast. \label{eq:error_bound_uncorrected} 
    \end{align}
    This is the reason why in panel e) of the sequence-describing Figures, e.g. \Cref{fig:seq_SLIC}, the smallest error strength that we consider 
    is $A/10$: no sequence (except by lowering $A_\ast$) is susceptible to errors much weaker than the effective interaction.

    Let us now assume that, instead, the sequence under consideration suppresses the error in first-order AHT, that is $H_{eff,err}=0$. Then \cite{brinkmannIntroductionAverageHamiltonian2016} AHT, or to be more precise the Magnus expansion, tells us that corrections to $H_{eff}$ are bounded by $||H^{rot}(t)||^2T/2$. For simplicity, we assume that this full contribution is reached through a single error term $H_{err}$. Now, the sufficient condition for a successful transfer
    interaction is given by 
    \begin{align} \label{eq:error_bound_AHT_order1}
    ||H_{err}||^2 \frac{T}{2}\ll A_\ast \Leftrightarrow ||H_{err}||^2 \ll \frac{2 A_\ast}{T} \sim \frac{JA_\ast}{\pi} 
    \end{align}
    where in the last step we inserted $T=2\pi/J$ as a representative time scale for polarisation sequences.
    This is the reason why we can expect sequences which suppress e.g. the dipolar field terms to first order to show signs of decay for $\Delta_{DF}$ somewhere between $A_\ast$ and $J$. Similarly, sequences with second order suppression will push more strongly towards the AHT-intrinsic limit of $||H_{err}||T\le 1$.

    The sequences which we regard in \Cref{sec:DFsuppressing_(2)} and \Cref{sec:2ch_seqs_theory} instead use a quicker time scale $\tilde{T}\sim 2\pi/\Omega$ in order to suppress the influence of dipolar fields, and thus are able to reach the stronger $||H_{err}||^2\ll 4\pi\ \Omega A_\ast$ for this kind of error.

    For sequences with $T\sim 2\pi/J$, the described approach to AHT does not predict the strong robustness to pulse errors $\Delta_0,\Delta_1$ for sequences like PulsePol which evidently show a scaling with $\Omega$ instead of $J$. For the case of pulse errors, suitable stronger bounds can be established by already including the errors as part of the rotating frame and then analysing their influence on $H_{eff}$ and $U(T)$ from this perspective. A corresponding analysis can be found in the supplementary material of \cite{korzeczekUnifiedPicturePolarization2023}.
    
	  \section{Sequences that are not adjusted for dipolar fields}\label{sec:unadjusted}

We will now discuss why and how polarisation sequences which do not actively account for the dipolar field will generally be susceptible to its influence and thus restrict the viable sample concentration. According to Average Hamiltonian theory Eq. \eqref{eq:H_eff_AHT}, the rotating frame interaction Eq. \eqref{eq:H_rot} is averaged over a period $T$ over which $S_z^{rot}(t)$ is controlled by the control sequence.
As the intended transfer Hamiltonian Eq. \eqref{eq:H_eff_transfer} relies solely on contributions to $S_x$ and $S_y$, well-designed polarisation sequences naturally avoid $S_z$ contributions to $S_z^{rot}(t)$ to maximize their transfer speed $A_\ast$.

Furthermore, a successful magnetisation build-up following Eq. \eqref{eq:H_eff_transfer} implies that in the rotating frame only $\expval{S_z}$ gives non-vanishing contributions. Thus, to a good approximation, usual sequences will yield
\begin{align} \label{eq:H_DFrot_unadjusted}
H^{rot}_{DF}(t) &= \Delta_{DF} \left( 3 \expval{S^{rot}_z(t)} S^{rot}_z(t) - \expval{\vec{S}^{rot}(t)}\cdot\vec{S}^{rot}(t) \right) \\ \nonumber
  &\approx -\Delta_{DF} \expval{\vec{S}^{rot}(t)}\cdot\vec{S}^{rot}(t) \\ \nonumber
  & = -\Delta_{DF} \expval{\vec{S}}\cdot\vec{S}\\ \nonumber
  & \approx -\Delta_{DF} \expval{S_z} S_z,
\end{align}
where we make use of the fact that the isotropic part of the effective Hamiltonian is invariant under rotations and hence stationary justifying the step from line two to line three. Only the isotropic term of the coupling contributes with a $z$-rotation proportional to the current magnetisation. Owing to its time independence, this contribution is not relevantly affected by the time average from Average Hamiltonian Theory
and thus directly contributes to $H^{rot}_{eff}$ of the sequence. This term does not leave the transfer Hamiltonian invariant and, by averaging it, suppresses the intended transfer interaction. As a result, there is a direct competition between $\Delta_{DF}$ and $A_\ast$.

\subsection*{Spin-Lock Induced Crossing (SLIC)} \label{sec:seq_SLIC}
\begin{figure}
		\centering
		\includegraphics[width=0.7\linewidth]{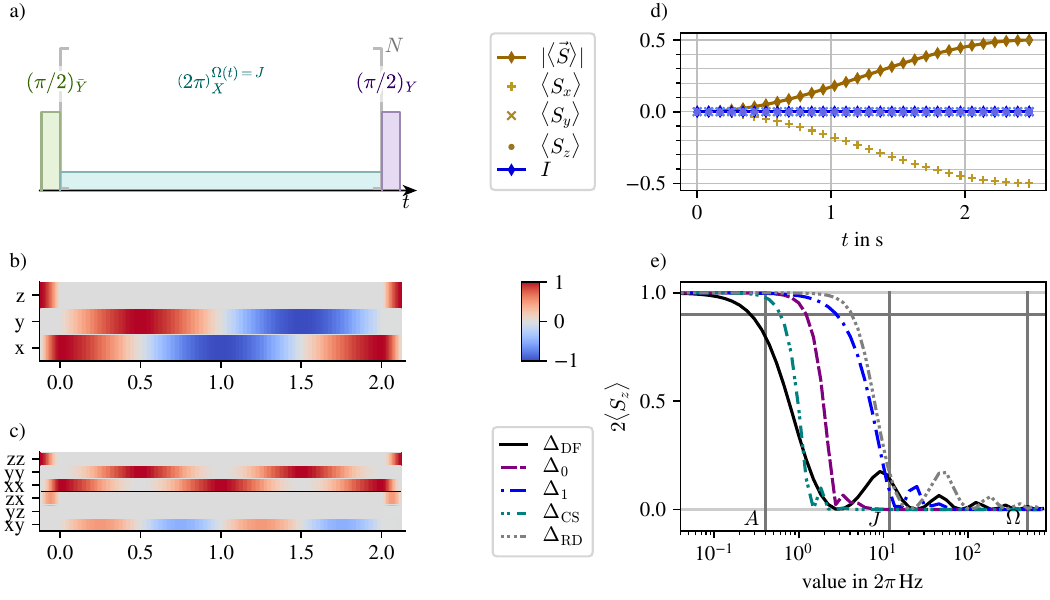}
        \caption{Characterisation of the SLIC sequence as described in \Cref{sec:seq_describing_figs}.
        a--c) use $\Omega=(2\pi)\,50\,$Hz for better readability.
        }
		\label{fig:seq_SLIC}
	\end{figure}

	The SLIC (Spin-Lock Induced Crossing \cite{PhysRevLett.111.173002}) sequence is a continuous-wave scheme for polarisation transfer which can be viewed as the most direct approach to creating transfer. Viewed as a pulsed sequences, the SLIC sequence can be defined as
    \begin{align}\label{eq:seq_SLIC}
    \pulse{\pi/2}{\bar{Y}} \left[ \pulse{2\pi}{X}^{\Omega(t)=J} \right]^N \pulse{\pi/2}{Y}.
    \end{align}
    Here, the continuous wave pulse with amplitude matching $\Omega(t) = J$ is described as an $N$-fold repetition of a $2\pi$-pulse, each with a duration $T=2\pi/J$. SLIC reaches the ideal transfer speed with $A_\ast =A$ because
    \begin{align}
        S_z^{rot}(t)&=S_x \cos Jt + S_y\sin Jt \label{eq:S_z_rot_SLIC}\, ,\\
        H_{eff}^{rot} &\approx \dfrac{1}{T}\int_0^{T}\mathrm{d}t\ A S_z^{rot}(t)(I_x\cos Jt - I_y\sin Jt) +H_{err}^{rot}(t) \\ \nonumber
        &= A\dfrac{S_x I_x - S_y I_y}{2} + \dfrac{1}{T}\int_0^T \mathrm{d}t\ H^{rot}_{err}(t)
    \end{align}
    which, except for the $H_{err}$ contribution, has the form of Eq. \eqref{eq:H_eff_transfer} with $A_\ast=A$, $S_{\tilde{x}}=S_x$ and $S_{\tilde{y}}=S_y$. Figure \ref{fig:seq_SLIC} shows a graphical representation of the pulses, $S_z^{rot}(t)$ and the properties of SLIC. Panel d) shows that in the frame of $H$ the magnetisation on $S$ accumulates in phase ($S_x$) with the continuous wave pulse without being reoriented over this period. Panel e) shows that as expected, already $\Delta_{DF}=A$ leads to a significant reduction of the transferred polarisation. 
    
    Note that pulse amplitude errors of magnitude $\Delta_1$ as described by Eq. \eqref{eq:H_err} add the error to the maximum amplitude available to the sequence $\Omega$ instead of the amplitude used in practice. Thus, the continuous wave amplitude is affected only by a fraction of the full error $\Delta_1$, namely $\Delta_1 J/\Omega$ which is an important detail for correctly interpreting the robustness of SLIC to amplitude errors $\Delta_1$ as shown in \cref{fig:seq_SLIC} and \cref{tab:seq_overview}.
    
	\subsection*{PulsePol} \label{sec:seq_PulsePol}
    PulsePol \cite{schwartzRobustOpticalPolarization2018} is a pulsed polarisation sequence that displays strong robustness to pulse errors $\Delta_0$ and $\Delta_1$. Due to its usefulness, we directly introduce generalised PulsePol \cite{tratzmillerParallelSelectiveNuclearspin2021,tratzmillerPulsedControlMethods} which has a phase $\varphi$ as free parameter and corresponds to the original PulsePol for $\varphi=\pm \pi/2$.
    
    The sequence can be defined as
    \begin{align*}
    \left[ \pulse{\pi/2}{X} \wait{\tau/4} \pulse{\pi}{Y} \wait{\tau/4} \pulse{\pi/2}{X} \cdot \pulse{\pi/2}{X+\varphi} \wait{\tau/4} \pulse{\pi}{Y+\varphi} \wait{\tau/4} \pulse{\pi/2}{X+\varphi} \right]^N,
    \end{align*}
    The resonance condition for any given $\varphi$ that yields the desired polarisation transfer Hamiltonian Eq. \eqref{eq:H_eff_transfer} that accumulates magnetisation along $\pm \hat{e}_z$ is given by $$J \tau = 2n(2\pi) \mp 2\varphi$$ for any non-negative integers $n$. Assuming ideal pulses, the effective coupling strength is given by $$A_\ast = A\, \dfrac{\sin^2(J\tau/8)}{J\tau/8}$$ which reaches a maximum at $J\tau=3\pi$, $\varphi =\pm \pi/2$ corresponding to standard PulsePol \cite{schwartzRobustOpticalPolarization2018}. 
    For any given value of $\varphi$, generalised PulsePol has $T=\tau = (2n(2\pi)+2\varphi)/J$ and results in $\alpha=2\varphi$. Note that, "resonances" that approach $J\tau= 2\pi k$, $\varphi=0,\pi$  (with integer $k$) do not lead to transfer as the time scale needed to construct the transfer terms diverges and the sequence approaches a CPMG-like \cite{maudsleyModifiedCarrPurcellMeiboomGillSequence1986, gullionNewCompensatedCarrPurcell1990} sensing sequence.
    
\begin{figure}
		\centering
		\includegraphics[width=\linewidth]{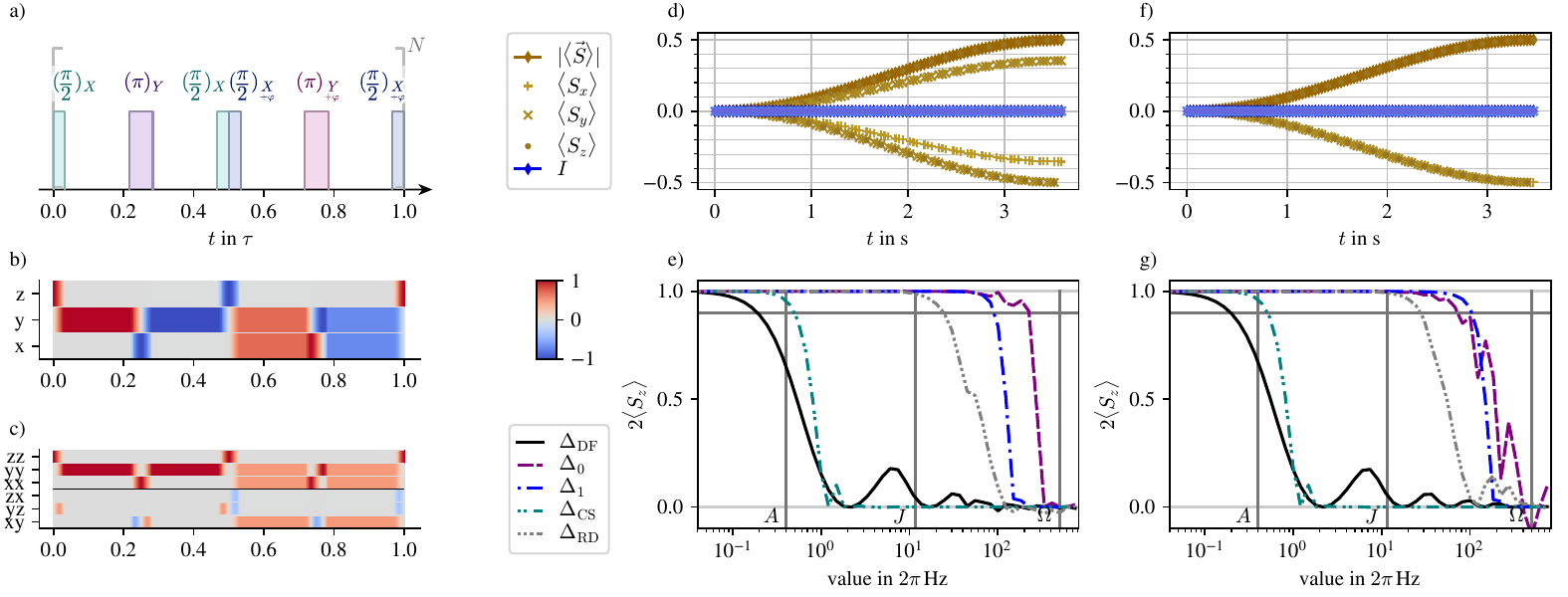}
        \caption{Characterisation of the PulsePol sequence for $\varphi=\pi/4$ and $J\tau= 3.5\pi$ as described in \Cref{sec:seq_describing_figs}. 
        a--c) use $\Omega=(2\pi)\,50\,$Hz for better readability.
        f) and g) correspond to d),e) but for generalized PulsePol with $\varphi=\pi/2$ and $J\tau=3\pi$}
		\label{fig:seq_PulsePol}
	\end{figure}

    As reference, we use $\varphi=\pi/4$ and $J\tau = 3.5\pi$. In Fig. \ref{fig:seq_PulsePol}e) we see that despite very good robustness to $\Delta_0$ and $\Delta_1$ errors, the susceptibility to dipolar fields is not improved compared to SLIC, and even slightly worse due to the lower effective coupling $A_\ast$. This matches our expectations from the discussion around Eq.~\eqref{eq:H_DFrot_unadjusted}.
    Compared to SLIC, PulsePol shows somewhat improved robustness to radiation damping $\Delta_{RD}$, while chemical shifts $\Delta_{CS}$ are unaffected by any sequence acting on $S$ and are thus slightly more problematic due to the lower $A_\ast$.
    Panels f,g) show the results for standard PulsePol with $\varphi=\pi/2$ and $J\tau=3\pi$ which mostly differs in its robustness to $\Delta_0$ errors which is strong, but somewhat weaker than for the generalised version.
    
	\clearpage
 
	\section{Dipolar field adjusted sequences}\label{sec:adjusted}
 
    The discussion of Eq. \eqref{eq:H_DFrot_unadjusted} showed that the leading order effect of the dipolar field in the co-rotating frame of a polarisation sequence can be captured by adding a Hamiltonian term of the form $H^{rot}_{DF}(t) \cong -\Delta_{DF} \expval{S_z} S_z$ which, for a known value of $\expval{S_z}$, simply amounts to a known additional rotation about the z-axis in the co-rotating frame. Such a rotation can be compensated for by adjusting the sequence such that it produces an equal but opposite effect.
   
   In order to identify the necessary modifications to a control sequence, let us assume the effective interaction of the modified sequence in the presence of $\Delta_{DF}$ to remain the ideal transfer of Eq. \eqref{eq:H_eff_transfer}. If we insert the resulting magnetisation buildup of the $S$ spin into $H_{DF}^{rot}(t)$ from Eq. \eqref{eq:H_DFrot_unadjusted} we arrive at
    \begin{align} \label{eq:dipfield}
       H_{DF}^{rot}(t) &\approx  - \dfrac{\Delta_{DF}}{2}\ \sin^2(\dfrac{A_\ast t}{4}) S_z.
    \end{align}

    For transfer to be successful in the presence of this additional term, the effective Hamiltonian $H^{rot}_{eff}$ needs to cancel
    it by changing the evolution of $S_z^{rot}(t)$. As $H_{DF}^{rot}(t)$ induces a precession of spin-$S$ in $H^{rot}$ about the $z$-axis, this can be achieved by introducing an opposing precession to $S_z^{rot}(t)$. The precise manner in which this is achieved will depend on the control sequence but common to all of them is that a resonance that was previously achieved with $J$ now needs to be changed to a time dependent resonance matching $J - \frac{\Delta_{DF}}{2} \sin^2(\frac{A_{\ast} t}{4})$ rather than $J$.
    In what follows we will discuss the required modifications for both continuous wave and for pulsed sequences.

    As adjusted sequences do not repeat, the AHT error bound estimate from Eq.~\ref{eq:error_bound_AHT_order1} can not be applied directly. Here, it is useful to treat the adjustments as a virtual error term which cancels (some of) the $\Delta_{DF}$ contributions and treat the unadjusted sequence as the basis for AHT.
 
	\subsection*{SLIC*} \label{sec:seq_SLIC*}
    Following the general reasoning above, we can amend SLIC to arrive at SLIC* simply by replacing the time independent resonance condition $\Omega(t) = J$ for the adjusted, now time-dependent, resonance condition $\Omega(t)=J - \frac{\Delta_{DF}}{2} \sin^2(\frac{A_{\ast} t}{4})$, such that
    \begin{align}\label{eq:seq_SLIC*}
    \pulse{\pi/2}{\bar{Y}} \left[ \pulse{CW}{X}^{\Omega(t)} \right] \pulse{\pi/2}{Y}
    \end{align}
    to ensure that Eq. \eqref{eq:dipfield} is compensated for.
    
\begin{figure}
		\centering
		\includegraphics[width=0.7\linewidth]{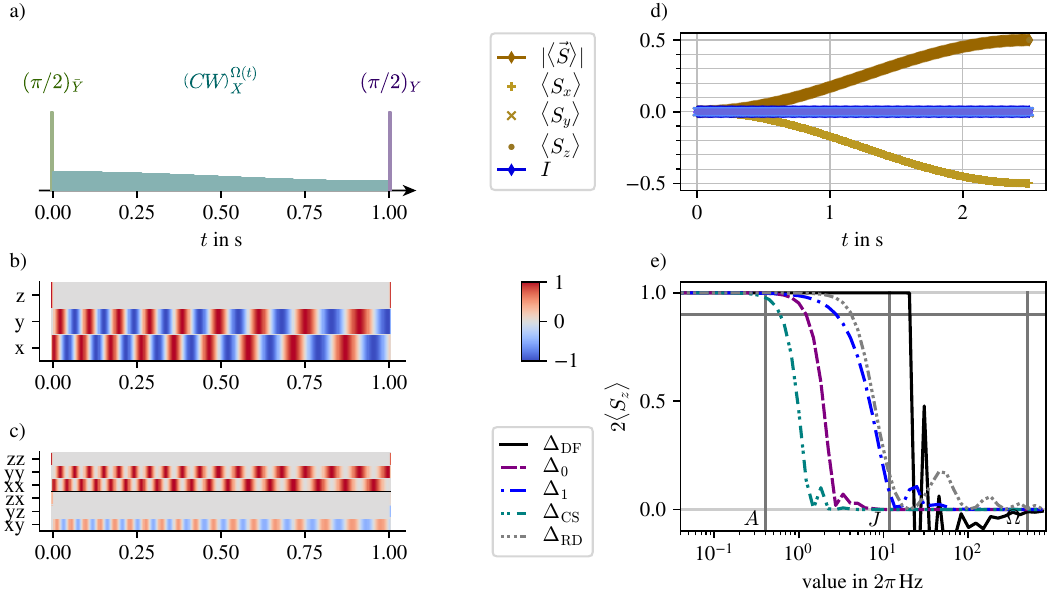}
        \caption{Characterisation of the SLIC* sequence for $\Delta_{DF}=J$ as described in \Cref{sec:seq_describing_figs}. 
        a--c) use $A=(2\pi)\,1\,$Hz and $\Omega=(2\pi)\,50\,$Hz for better readability.
        }
		\label{fig:seq_SLIC*}
	\end{figure}

    Fig. \ref{fig:seq_SLIC*} shows that this adjustment allows for successful transfer for significantly stronger, but known, dipolar fields than SLIC. While SLIC fails when $\Delta_{DF}\cong A_{\ast}$, SLIC* remains robust until $\Delta_{DF}\cong 2J$ which is significantly stronger than the limit expected from the worst-case second-order corrections to AHT.
    Given that the adjustment of SLIC* directly cancels $H_{DF}^{rot}(t)$ without relying on AHT, the pure $\Delta_{DF}$ error terms vanishes to all orders. The limitations can only come from how AHT constructs the effective transfer interaction $A_\ast$ and its higher order corrections which can still include $\Delta_{DF}$. The lowest-order mixing term can be bounded by $|A\Delta_{DF}|T/2\sim \pi |A\Delta_{DF}|/2J$, which is smaller than the intended interaction $A_\ast$ for $|\Delta_{DF}|\lesssim 2J/\pi$. This argument underestimates the numerically established limit close to $2J$ largely because it takes a worst-case estimate for the second-order error ignoring possible cancellations due to changing signs of the integrand. However, it shows how an estimate independent of the parameter $A$ can be reached. Remarkably, in \cref{sec:DFenabled} 
    we will see that moderate $\Delta_{DF}$ can not only be compensated, but even be advantageous for transfer. Before discussing that aspect, let us consider first how the dipolar field adjustment plays out in a pulsed sequence.

	\subsection*{PulsePol*} \label{sec:seq_PulsePol*}    
\begin{figure}
		\centering
		\includegraphics[width=0.7\linewidth]{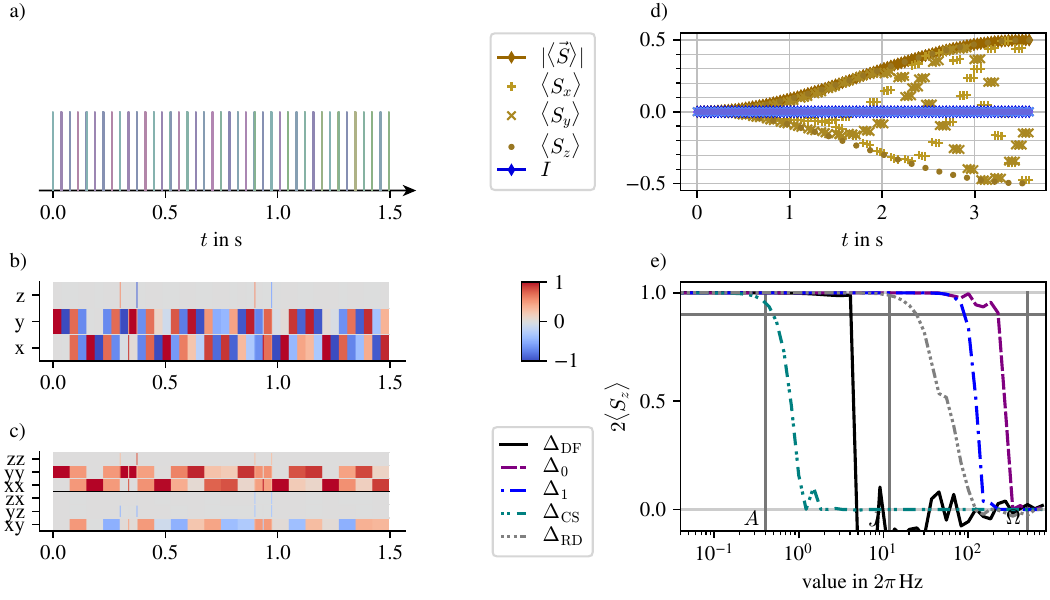}
        \caption{Characterisation of the PulsePol* sequence for $\varphi=\pi/4$ and $J\tau= 3.5\pi$ as described in \Cref{sec:seq_describing_figs}. 
        a--c) use $A=(2\pi)\,1\,$Hz and $\Omega=(2\pi)\,50\,$Hz for better readability.
        a--d) use $\Delta_{DF}=(2\pi)\,1\,$Hz.
        }
		\label{fig:seq_PulsePol*}
	\end{figure}

    In PulsePol the resonance condition is not governed by the strength of the Rabi frequency but by judicious choice of the pulse separations. We could use these to
    counter the effect of the dipolar field. However, adjusting the pulse separations also affects $A_\ast$ which would become time-dependent as a result so that Eq. \eqref{eq:dipfield} would require the change $A_\ast t\to \int_0^t\mathrm{d}t' A_\ast(t')$ and complicate a solution.
    Alternatively, here we make use of the fact that the generalised PulsePol sequence can compensate for the effect of the dipolar field by following the inverse of the rotation around the z-axis induced by Eq. \eqref{eq:dipfield}. To this end we need to determine the integrated effect of the dipolar field term over time. Indeed, Eq. \eqref{eq:H_DFrot_unadjusted} introduces the phase
    \begin{align*}
    \phi(t) &= \int_0^t \mathrm{d}t'\ \left[-\dfrac{\Delta_{DF}}{2} \ \sin^2(\frac{ A_\ast t}{4})\right]\\
        &= -\dfrac{\Delta_{DF}}{2} \left[ \dfrac{t}{2} - \dfrac{\sin(A_\ast t/2)}{A_\ast} \right]
    \end{align*}
    which describes the precession of 
    $S_z$.   
    Knowing these phases, we can choose $\phi_n := \phi(n\tau)$ to define PulsePol* as
    \begin{align*}
    \Pi_{n=1}^N \left[ \pulse{\pi/2}{X-\phi_n} \wait{\tau/4} \right. & \pulse{\pi}{Y-\phi_n} \wait{\tau/4} \pulse{\pi/2}{X-\phi_n}\cdot 
    \pulse{\pi/2}{X+\varphi-\phi_n} \wait{\tau/4} & \left. \pulse{\pi}{Y+\varphi-\phi_n} \wait{\tau/4} \pulse{\pi/2}{X+\varphi-\phi_n}\right].
    \end{align*}
    The sequence using $\varphi=\pi/4$ and $J\tau=3.5\pi$ is shown in \Cref{fig:seq_PulsePol*}. We see that its properties are essentially unchanged compared to normal PulsePol, while it can compensate small, known values of the dipolar field strength $\Delta_{DF}$. In this sequence, the adjustment compensates for the dynamics from $\Delta_{DF}$ after every $1\tau$, such that dipolar fields are first-order corrected on that time scale. With this, the magnitude of acceptable dipolar fields is well in line with the estimate from AHT which expects corrections to dominate as soon as $\Delta_{DF}\sim \sqrt{A_\ast J}$.

	\clearpage
 
	\section{DF-enabled}\label{sec:DFenabled}
    
	\subsection*{Amplitude swept SLIC} \label{sec:seq_amp_swept_SLIC}
In Eq. \eqref{eq:H_DFrot_unadjusted} we have seen that the dipolar field influence during the continuous-wave pulse of SLIC is equivalent to a modification of the pulse amplitude $\Omega(t)$ proportional to both the accumulated magnetisation $\expval{S_z}$ and $\Delta_{DF}$. 
One can now imagine a situation where $\Omega(t)$ starts off-resonant before magnetisation has accumulated and is continuously swept towards and across the resonance condition. 
In the absence of significant dipolar fields, this scheme has been used in e.g. \cite{marshallRadioFrequencySweepsMicrotesla2023a} where it allows to trade off the overall transfer rate $A_\ast$ with robustness to $B_1$ amplitude errors $\Delta_1$ \cite{korzeczekUnifiedPicturePolarization2023}.

\begin{figure}
		\centering
		\includegraphics[width=\linewidth]{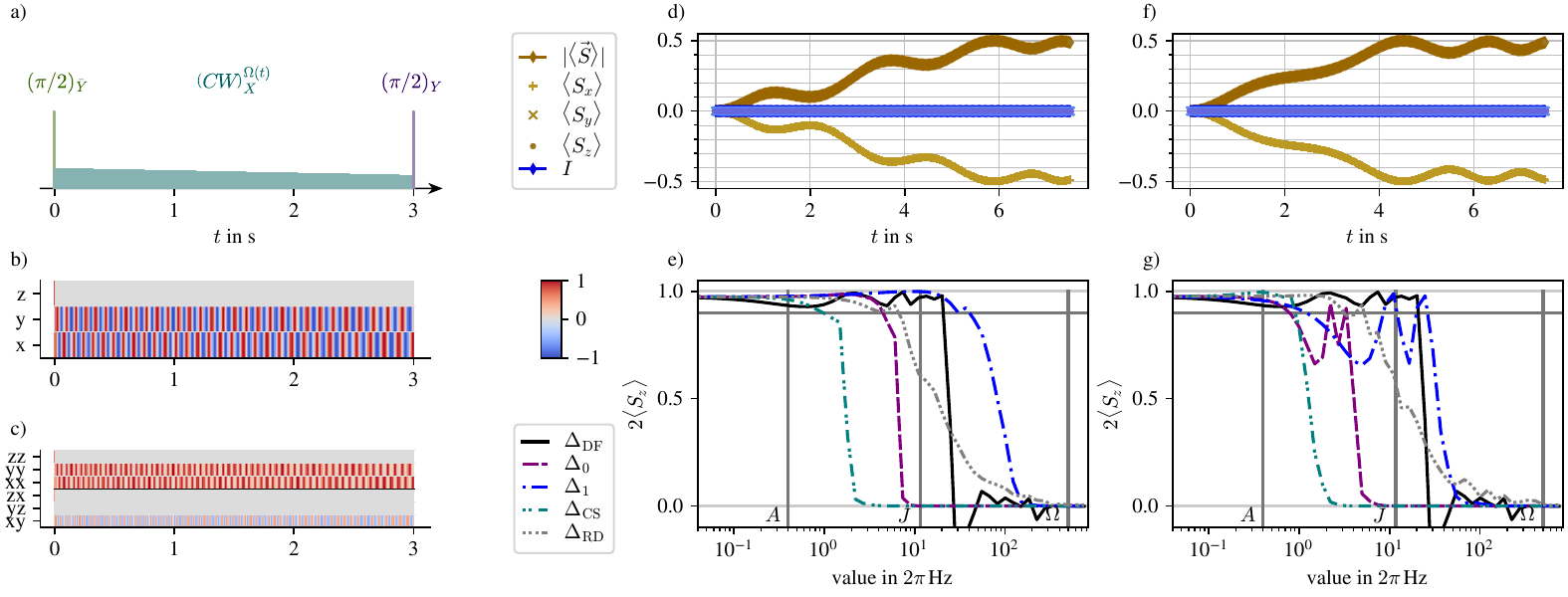}
        \caption{Characterisation of the amplitude swept SLIC sequence as described in \Cref{sec:seq_describing_figs}. 
        a--c) use $A=(2\pi)\,1\,$Hz and $\Omega=(2\pi)\,50\,$Hz for better readability.
        a--e) use $\Delta_{DF}=(2\pi)\,3\,$Hz except for the line in panel e where $\Delta_{DF}$ is varied.
        f,g) correspond to d,e) but without the distant dipolar field stabilising the dynamics, that is $\Delta_{DF}=0$.}
		\label{fig:seq_amp_swept_SLIC}
	\end{figure}

In the presence of significant dipolar fields however, i.e. approximately in the regime $A < |\Delta_{DF}| < 2J$, we find numerically that the dipolar field introduces an intriguing stabilising effect. Assuming a fixed value of $\Delta_{DF}$, during the sweep, when $\Omega(t)$ approaches $J$, magnetisation starts to accumulate and thus effectively shifts the resonance condition to $\Omega(t)=J - \Delta_{DF} \expval{S_z}/2$. We find that for opposite signs of $\dot \Omega(t)$ 
and $\Delta_{DF}\langle S_z\rangle$, this leads to a stabilizing dynamics where the magnetisation approximately follows 
\begin{align} \label{eq:equilibriumnonlinear}
\expval{S_z}&\approx \dfrac{J-\Omega(t)}{\Delta_{DF}/2},
\end{align}
after the resonance condition $\Omega(t)=J$ is first met and before $\langle S_z\rangle$ saturates due to $S$ being fully magnetised. Due to the actual magnetisation oscillating around this point, the transfer is not exact but remains successful for a range of sweep rates for $\Omega(t)$. Thanks to this property, the presence of a moderately high dipolar field allows for a relatively fast transfer rate $A_\ast$, while maintaining good robustness against amplitude errors (cf.~\Cref{fig:seq_amp_swept_SLIC}). As a reasonable, but not optimised, choice we use the following sequence 
\begin{align}\label{eq:seq_SLIC*}
    \pulse{\pi/2}{\bar{Y}} \left[ \pulse{CW}{X}^{\Omega(t)} \right] \pulse{\pi/2}{Y}
\end{align}
with a linear sweep defined by
\begin{align}
    \Omega(t)=J- \dfrac{\Delta_{DF}}{|\Delta_{DF}|}[-A+(|\Delta_{DF}|/2+3A)\dfrac{t}{2\pi/A_\ast}]
\end{align} 
where $A_\ast=\dfrac{1}{3}\ A$. This choice remains applicable at $\Delta_{DF}=0$ and ensures that the rate of change of the equilibrium condition Eq. \eqref{eq:equilibriumnonlinear} does not exceed the rate of polarisation transfer, allowing the system to remain near equilibrium throughout the process. \Cref{fig:seq_amp_swept_SLIC} e) shows the robustness to errors when a distant dipolar field of $\Delta_{DF}=(2\pi)\ 3\,$Hz is included as part of the intended Hamiltonian, with additional single-parameter errors applied (except for the $\Delta_{DF}$ line, which assumes perfect control parameters). Amplitude errors affect the dynamics when $J$ falls outside the error-adjusted sweep range, preventing magnetisation from building up from the start. Detuning errors remain suppressed up to approximately the pulse amplitude, which is limited by $J$. For comparison, panels f,g) show the results when the sweep is not supported by distant dipolar fields and $\Delta_{DF}=(2\pi)\ 0\,$Hz is used, for example, for evaluating pulse error robustness.

Similarly to the sequences SLIC* and PulsePol* described in \cref{sec:adjusted}, the amplitude-swept SLIC sequence does not suppress dipolar field terms and therefore relies on the mean-field approximation of the distant dipolar field. As with SLIC* and PulsePol*, amplitude-swept SLIC is adapted to a specific dipolar field strength $\Delta_{DF}$. Consequently, the geometry of the sample and the spatial inhomogeneity of $\Delta_{DF}(\vec{r})$ can significantly impact the stability - and thus the practical feasibility - of amplitude-swept SLIC (cf. \cref{sec:DDF}).

	\clearpage
	\section{DF-suppressing (1)}\label{sec:DFsuppressing_(1)}

	While the sequences introduced in sections \Cref{sec:adjusted} and \Cref{sec:DFenabled} allow for successful transfer even with significant strengths of $\Delta_{DF}$, they achieve this by adapting to, or even making use of the distant dipolar field. An alternative approach to ensure successful polarisation transfer suppresses the dipolar interaction altogether:

Control sequences are termed dipolar field suppressing sequences when they satisfy the property 
\begin{align}\label{eq:DF_suppressing}
    H_{DF,eff}^{rot} &= \dfrac{1}{\tilde{T}}\int_0^{\tilde{T}}\mathrm{d}t \ H^{rot}_{DF}(t) = 0,
\end{align}
where $\tilde{T}$ is closely related to the periodicity $T$ of $H_{ctrl}(t)$, but might not be identical. As long as $|\Delta_{DF}\tilde{T}|\ll 1$, dipolar field suppressing sequences do not rely on the mean-field or isotropic approximations for Eq. \eqref{eq:delta_DF} and thus provide a relatively assumption-free method for successful polarisation transfer at high concentrations where unadjusted sequences (\cref{sec:unadjusted}) are no longer successful.
In this section, we present sequences which tie the dipolar field suppression time scale $\tilde{T}$ to the resonance frequency $J$ which allows us to use adaptations of existing polarisation sequences. In \cref{sec:DFsuppressing_(2)}, we introduce sequences which use a $\tilde{T}$ that is only limited by the strength of the available control field $\Omega$.

\subsection*{MA-SLIC} \label{sec:seq_MASLIC}
    Lee-Goldburg decoupling \cite{goldburgNuclearMagneticResonance1963,leeNuclearMagneticResonanceLineNarrowing1965} is a well-established method for achieving dipolar-field suppression Eq. \eqref{eq:DF_suppressing}. A well-chosen off-resonant continuous-wave pulse induces an effective rotation around an axis that differs from $\vec{B}_0$ by the magic angle $\alpha\ind{MA}=\arccos(\sqrt{1/3})$ and thus averages the dipolar field contribution $H_{DF}^{rot}(t)$ to zero over one period of oscillation.
    Magic Angle SLIC as a sequence is a variant of SLIC which uses this magic angle detuned continuous wave pulse at an amplitude that is suited to inducing a resonance while also suppressing the dipolar field. In our pulsed notation, we can write
    
    \begin{align*}
    \pulse{\alpha\ind{MA}}{\bar{Y}} \left[ \pulse{2\pi}{X;\Delta=\sqrt{1/3} J}^{\Omega=\sqrt{2/3}J} \right]^N \pulse{\alpha\ind{MA}}{Y}.
    \end{align*}

\begin{figure}
		\centering
		\includegraphics[width=0.7\linewidth]{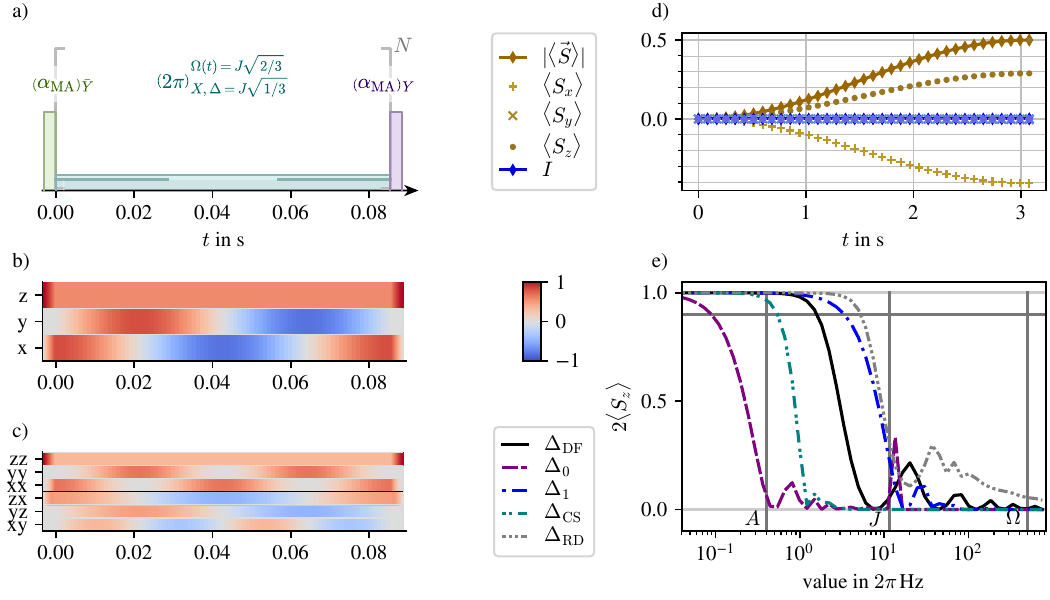}
        \caption{Characterisation of the MA-SLIC as described in \Cref{sec:seq_describing_figs}. 
        a--c) use $\Omega=(2\pi)\,50\,$Hz for better readability.
        }
		\label{fig:seq_MASLIC}
	\end{figure}

    Where the initial and final pulses are correspondingly adjusted to reflect the new axis of rotation. A fully equivalent sequence has been proposed recently in the context of DNP under the name Magic-NOVEL \cite{javedMagicNOVELSuppressingElectron2025}.
    Compared to SLIC with $A_\ast=A$, the effective coupling is reduced to $A_\ast=\sqrt{2/3}\,A$ and \Cref{fig:seq_MASLIC} shows the corresponding properties. As we would expect, panel e) shows an improved limit to the acceptable $\Delta_{DF}$ that is in line with the AHT estimate of $\Delta_{DF}\sim \sqrt{J A_\ast /\pi}$ from Eq. \eqref{eq:error_bound_AHT_order1}, however the off-resonant drive is more susceptible to pure detuning errors $\Delta_0$ than SLIC.
    With this, MA-SLIC demonstrates the viability of combining dipolar decoupling Eq. \eqref{eq:DF_suppressing} with a polarisation sequence Eq. \eqref{eq:H_eff_transfer}. Next, we will introduce suitable adaptations of PulsePol, which allow for a high robustness to pulse errors $\Delta_{0/1}$ in addition to dipolar decoupling. 
    
\subsection*{MA-PulsePol} \label{sec:seq_MAPulsePol}

\begin{figure}
		\centering
		\includegraphics[width=1\linewidth]{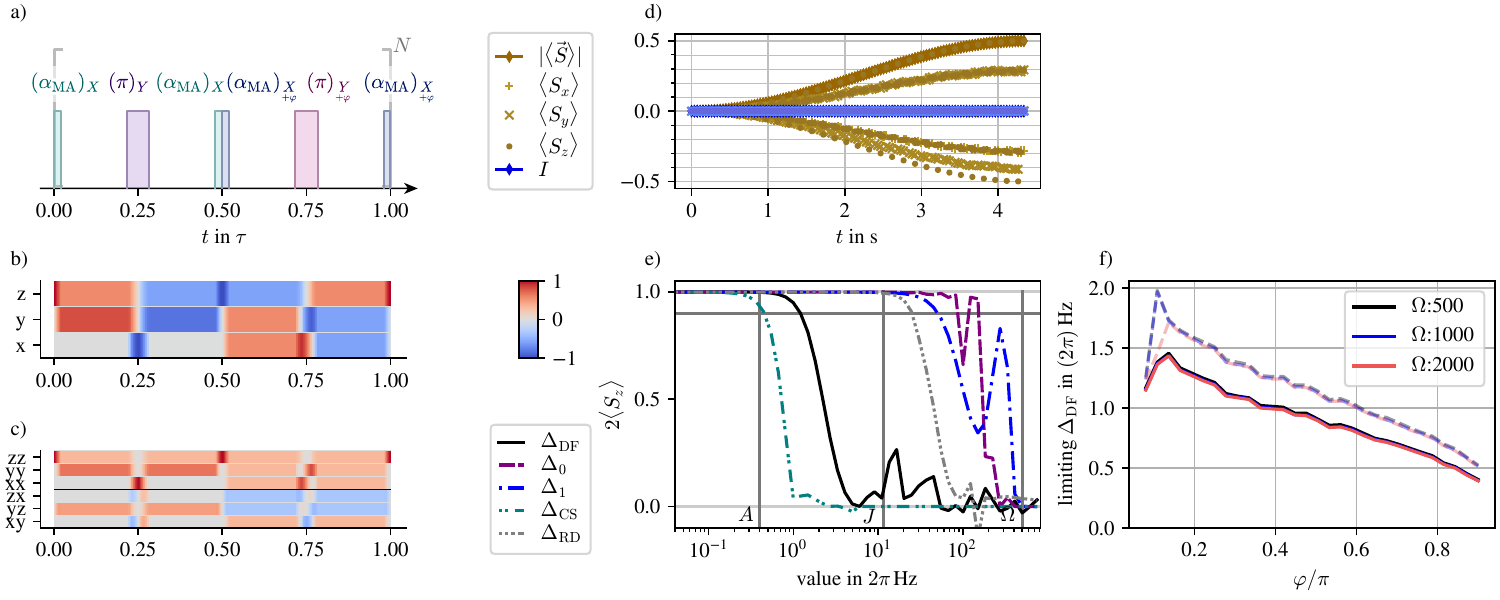}
        \caption{Characterisation of the MA-PulsePol sequence with $\varphi=\pi/4$ and $J\tau=3.5\pi$ as described in \Cref{sec:seq_describing_figs}. 
        a--c) use $\Omega=(2\pi)\,50\,$Hz for better readability.
        f) shows the limiting dipolar field $\Delta_{DF}$ for a variety of phases $\varphi$ using $J\tau=4\pi-2\varphi$ and different available Rabi amplitudes $\Omega$. The solid (dashed) lines correspond to the limit where $90\,$\% ($80\,$\%) of full transfer are reached.}
		\label{fig:seq_MAPulsePol}
	\end{figure}

    Magic-Angle PulsePol addresses the dipolar field challenge by turning the $\pi/2$ pulses into $\alpha\ind{MA}$ pulses, giving
    \begin{align*}
    \left[ \pulse{\alpha\ind{MA}}{X} \wait{\tau/4} \right. & \pulse{\pi}{Y} \wait{\tau/4} \pulse{\alpha\ind{MA}}{X} \cdot  \pulse{\alpha\ind{MA}}{X+\varphi} \wait{\tau/4}  \left. \pulse{\pi}{Y+\varphi} \wait{\tau/4} \pulse{\alpha\ind{MA}}{X+\varphi} \right]^N \, .
    \end{align*}
    This choice ensures that during waiting times $S_z^{rot}(t)$ always includes a component $\pm\sqrt{1/3}S_z$ while the $x$-$y$ components with total amplitude $\sqrt{2/3}$ follow the usual PulsePol precession and thus eventually ensure dipolar decoupling on a time scale related to $J$ and depending on $\varphi$.
    The resonance condition as a function of $\varphi$ remains $J\tau = 2n (2\pi) \mp 2\varphi$ and assuming ideal pulses, the effective coupling strength becomes $A_\ast=A_\perp \sqrt{\dfrac{2}{3}} \dfrac{\sin^2(J\tau/8)}{J\tau/8}$, that is approximately $18\,\%$ less than for PulsePol. 
    Thus, the penalty to transfer speed is the same as that observed when moving from SLIC to MA-SLIC. As shown in \Cref{fig:seq_MAPulsePol}, the pulse error robustness remains close to that of the original PulsePol (cf.~\Cref{fig:seq_PulsePol})). Panel f) illustrates that dipolar field suppression benefits from smaller values of $\varphi$, which leads us to choose $\varphi=\pi/4$ with $J\tau=3.5\pi$ as the reference MA-PulsePol, which corresponds to $\alpha=2\varphi=\pi/2$ (cf.~\cref{eq:Urot_precession}). The resulting acceptable dipolar field $\Delta_{DF}$ fits expectations based on first-order suppression on the $J$ time scale.
    
\subsection*{DF-PulsePol} \label{sec:seq_DF_PulsePol}
\begin{figure}
		\centering
		\includegraphics[width=1\linewidth]{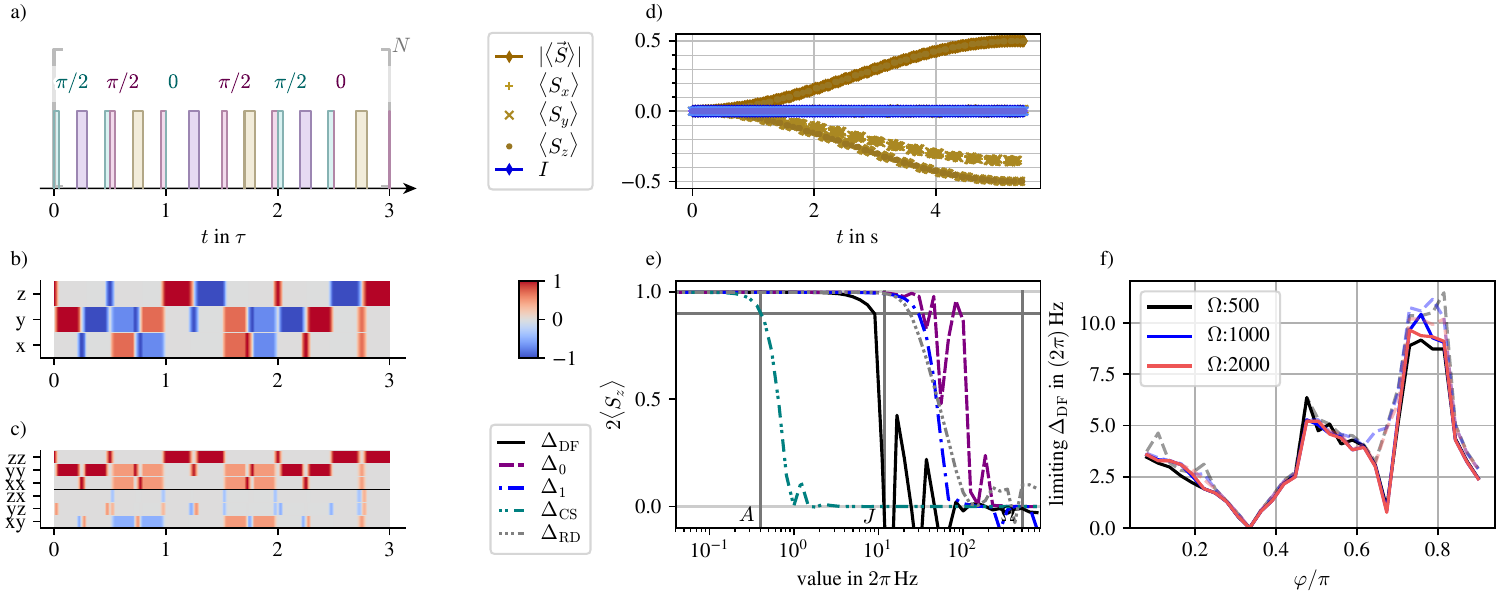}
        \caption{Characterisation of the DF-PulsePol sequence with $\varphi=3\pi/4$ and $J\tau=2.5\pi$ as described in \Cref{sec:seq_describing_figs}. 
        a--c) use $\Omega=(2\pi)\,50\,$Hz for better readability.
        f) shows the limiting dipolar field $\Delta_{DF}$ for a variety of phases $\varphi$ using $J\tau=4\pi-2\varphi$ and different available Rabi amplitudes $\Omega$. The solid (dashed) lines correspond to the limit where $90\,$\% ($80\,$\%) of full transfer are reached.}
		\label{fig:seq_DF_PulsePol}
	\end{figure}

    At the cost of slightly weakening the pulse error robustness $\Delta_{0/1}$ and effective coupling $A_\ast$, it is possible to significantly improve on the dipolar field suppression of MA-PulsePol:
    Dipolar field PulsePol (DF-PulsePol) is inspired by a similar adaptation to PulsePol which was developed in \cite{tratzmillerPulsedControlMethods}. 
    Compared to PulsePol, we remove the $\pi/2$ pulses adjacent to every third $\pi$ pulse to introduce waiting times with $S_z^{rot}(t)=\pm S_z$.
    With this, the sequence repeats after $3\tau$ and is defined by

    \begin{align*}
    \left[ \pulse{\pi/2}{X} \right. \wait{\tau/4} \pulse{\pi}{Y} \wait{\tau/4} \pulse{\pi/2}{X} & \pulse{\pi/2}{X+\varphi} \wait{\tau/4} \pulse{\pi}{Y+\varphi} \wait{\tau/4} \pulse{\pi/2}{X+\varphi} \\ \nonumber 
     \wait[4.4em]{\tau/4} \pulse{\pi}{Y} \wait[4.4em]{\tau/4}  & \pulse{\pi/2}{X+\varphi} \wait{\tau/4} \pulse{\pi}{Y+\varphi} \wait{\tau/4} \pulse{\pi/2}{X+\varphi} \\ \nonumber
    \pulse{\pi/2}{X}\wait{\tau/4} \pulse{\pi}{Y} \wait{\tau/4} \pulse{\pi/2}{X}& \wait[5.4em]{\tau/4} \left. \pulse{\pi}{Y+\varphi} \wait[5.4em]{\tau/4}  \right]^N\, .
    \end{align*}
    To emphasise the positions where PulsePol would include $\pi/2$ pulses, we have drawn longer lines to indicate the corresponding waiting times. These also have a slightly higher duration $t_{\text{wait}}=\tau/4-\pi/(2\Omega)$ as only one half of the neighbouring $\pi$ pulse needs to be fitted into the $\tau_{\text{wait}}=\tau/4$ time span.
    The resonance condition for $\pm \hat{e}_z$ magnetisation for any given $\varphi$ remains $J\tau = 2n (2\pi) \mp 2\varphi$ and assuming ideal pulses, the effective coupling strength is now given by $A_\ast=A_\perp \dfrac{2}{3} \dfrac{\sin^2(J\tau/8)}{J\tau/8}$, i.e. a $1/3$ reduction compared to PulsePol. Each sequence repetition creates a net precession of $\alpha=6\varphi$. As the quality of dipolar field suppression strongly depends on $\varphi$ (cf.~\Cref{fig:seq_DF_PulsePol}f) , we choose $\varphi=3\pi/4$ with $J\tau=2.5\pi$.
    
    In \Cref{fig:seq_DF_PulsePol}, we see that DF-PulsePol handles $\Delta_{DF}$ values close to $J$ well, while showing significant pulse error robustness. Even though the latter is not as effective as for original PulsePol, it is not limited by molecule inherent properties like $J$. Therefore pulse error robustness can be increased by stronger control fields or the use of composite pulses.
    The robustness to dipolar fields is stronger than first order suppression, and for the choice of $\varphi\approx 3\pi/4$ even second order suppression, can account for. A plausible argument for this might be that under ideal and instantaneous pulses, even when considering the effect of $H_{DF}^{rot}(t)$ directly, DF-PulsePol either places $S_z^{rot}(t)$ in the $x$-$y$-plane -- similar to unadjusted sequences, where, according to Eq. \eqref{eq:H_DFrot_unadjusted}, the interaction approximately corresponds to a pure resonance shift -- or fully along the $z$-axis, in which case its effect is again a pure, but opposite, $z$-precession. Thus, the main effect of $\Delta_{DF}$ 
    will be this accumulation and reversal of the additional phase over every $3\tau/2$ interval. This particular effect can be expected to lead to distortions to $H_{eff}^{rot}$ but will not add a pure $S$-spin precession. 
    Although this argument makes the relatively strong robustness to dipolar fields plausible, a more detailed treatment would be necessary to verify the mechanism, and to explain details of the strong $\varphi$-dependence that we find in \Cref{fig:seq_DF_PulsePol}f).
    
    \subsection*{M2A-PulsePol} \label{sec:seq_M2A_PP}
    
    While DF-PulsePol is as good at handling dipolar fields as one can expect from a sequence with its time-scale matched to the $J$-precession, and MA-PulsePol keeps more of the transfer speed and pulse-error robustness of PulsePol, there are situations where a compromise between these properties is of interest.

    An example for such a compromise is two-angle MA-PulsePol (here abbreviated as M2A-PulsePol). Instead of using the fixed $\alpha\ind{MA}$ rotation angle for all "$\pi/2$" pulses as in MA-PulsePol, or the fully removed $\pi/2$ pulses of DF-PulsePol, M2A-PulsePol suitably alternates between $\pi/4$ and $\pi/2$ for the rotation angle. This allows for avoiding "odd" rotation angles, improving dipolar field suppression compared to MA-PulsePol thanks to a more time-symmetric $S_z^{rot}(t)$, and mostly retaining both the transfer speed and pulse-error robustness of the latter. M2A-PulsePol is defined by
    \begin{align*}
    \left[ \pulse{\pi/4}{X} \right. \wait[2em]{\tau/4} \pulse{\pi}{Y} \wait[2em]{\tau/4} \pulse{\pi/4}{X} &\cdot \pulse{\pi/2}{X+\varphi} \wait{\tau/4} \pulse{\pi}{Y+\varphi} \wait{\tau/4} \pulse{\pi/2}{X+\varphi}\cdot \\ \nonumber 
     \pulse{\pi/4}{X} \wait[2em]{\tau/4} \pulse{\pi}{Y} \wait[2em]{\tau/4} &\pulse{\pi/4}{X} \cdot \pulse{\pi/4}{X+\varphi} \wait[2em]{\tau/4} \pulse{\pi}{Y+\varphi} \wait[2em]{\tau/4} \pulse{\pi/4}{X+\varphi}\cdot \\ \nonumber \pulse{\pi/2}{X} &\wait{\tau/4} \pulse{\pi}{Y} \wait{\tau/4} \pulse{\pi/2}{X} \cdot \pulse{\pi/4}{X+\varphi} \wait[2em]{\tau/4} \pulse{\pi}{Y+\varphi} \wait[2em]{\tau/4} \left. \pulse{\pi/4}{X+\varphi}  \right]^N.
    \end{align*}
    Similar to DF PulsePol, we have drawn the waiting times adjacent to $\pi/4$ pulses slightly longer to indicate that only a total pulse duration of $3\pi/(4\Omega)$ needs to be subtracted from their $\tau_{\text{wait}}=\tau/4$.  
    The resonance condition remains that of PulsePol, with the effective coupling $A_\ast=A_\perp \dfrac{1+2\sqrt{1/2}}{3} \dfrac{\sin^2(J\tau/8)}{J\tau/8}$, that is approximately $20\,\%$ less than for PulsePol. Each sequence repetition creates a net precession of $\alpha=6\varphi$.
    
\begin{figure}
		\centering
		\includegraphics[width=\linewidth]{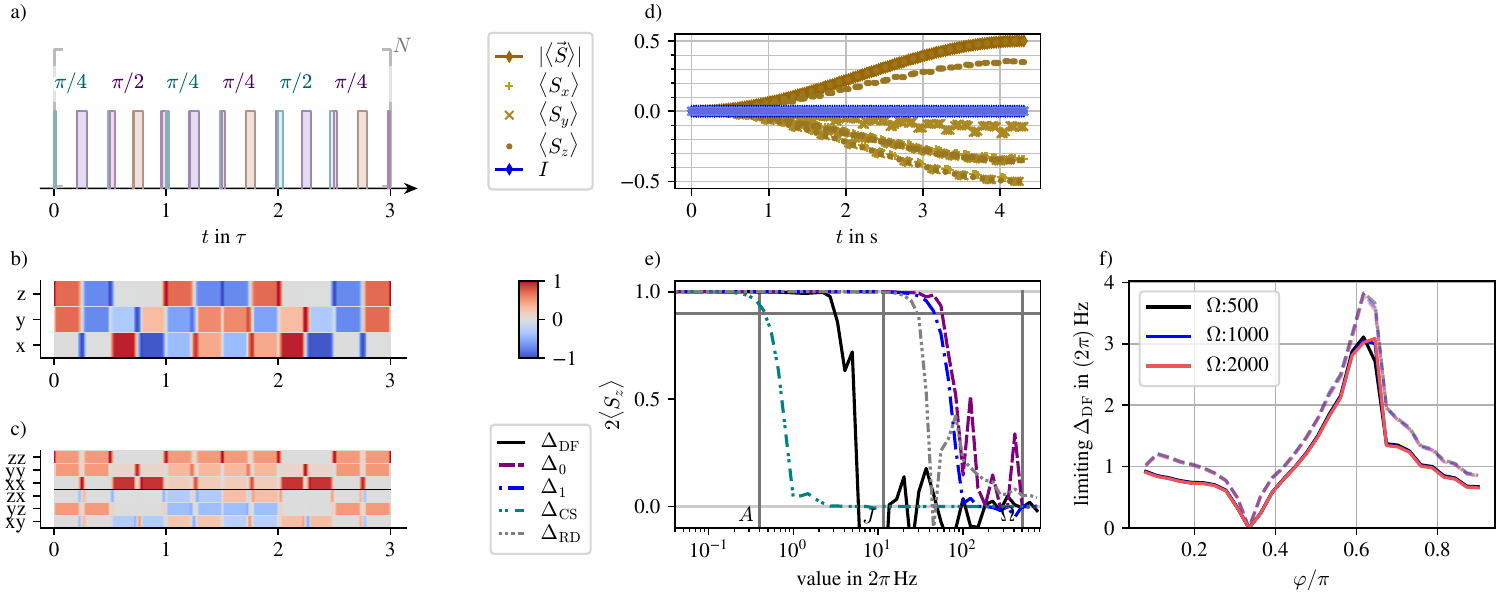}
        \caption{Characterisation of the M2A-PulsePol sequence with $\varphi=0.6\pi$ and $J\tau=2.8\pi$ as described in \Cref{sec:seq_describing_figs}. 
        a--c) use $\Omega=(2\pi)\,50\,$Hz for better readability.
        f) shows the limiting dipolar field $\Delta_{DF}$ for a variety of phases $\varphi$ using $J\tau=4\pi-2\varphi$ and different available Rabi amplitudes $\Omega$. The solid (dashed) lines correspond to the limit where $90\,$\% ($80\,$\%) of full transfer are reached.}
		\label{fig:seq_M2APulsePol}
	\end{figure}

    In \Cref{fig:seq_M2APulsePol} and \Cref{tab:seq_overview} we see that the properties of M2A-PulsePol indeed lie between those of MA-PulsePol and DF-PulsePol when choosing $\varphi=0.6\pi$ and $J\tau=2.8\pi$. Thanks to its approximately time-symmetric quadratic $S_z^{rot}(t)$ contributions (\Cref{fig:seq_M2APulsePol}c)) it suppresses most of the second order dipolar field contributions, while the more even "$\pi/2$" pulse durations retain more of the symmetry of the original PulsePol and thus pulse error robustness.
    
	\clearpage
	\section{DF-suppressing (2)}\label{sec:DFsuppressing_(2)}
  
	In this section we present sequences which average $H_{DF}^{rot}(t)$ to zero on a time scale $\tilde{T}$ that is independent of the resonance condition given by $J$ and only limited by the available Rabi amplitude $\Omega$.

For finding suitable sequences, we can use existing dipolar decoupling sequences from the solid-state NMR literature \cite{burumLowpowerMultipulseLine1981,fungComparisonHomonuclearDipolardecoupling1987} where suppressing the line-broadening from homonuclear dipolar coupling is of interest, while the $\tilde{T}$-averaging should ideally retain most of the amplitude of $S_z^{rot}(t)$ to retain the information of heteronuclear couplings in the NMR spectrum. 
In order to create a polarisation sequence, we can modify the dipolar decoupling sequences to add a comparatively slow precession to $S_z^{rot}(t)$ which can now be resonant to $J$ and allow for transfer.
In principle, the added precession can interfere with the dipolar decoupling: As an estimate, we can use Average Hamiltonian Theory for the unmodified sequence and regard the added precession as an "error term" of magnitude $J$. In this picture any resulting change to $H_{eff}^{rot}$ with terms proportional to $\Delta_{DF}$ must arise from second or higher order corrections with $J\tilde{T} \Delta_{DF}$ as factors. Given that $\Delta_{DF}\tilde{T}\ll 1$ is assumed by AHT already, this gives a reason to expect that the polarising versions of dipolar decoupling sequences will retain at least a good fraction of the decoupling properties.

The addition of a $J$-scale overall precession to achieve a polarising sequence can be applied to all dipolar decoupling sequences which do not average $S_z^{rot}(t)$ to zero on the fast $\tilde{T}$ time scale.
We choose the comparably simple Lee-Goldburg continuous-wave, and BLEW-12 as well as MREV-8 pulsed dipolar decoupling sequences as examples for which we present suitably adapted polarising versions. As these sequences directly follow a $J$-scale suppression for the $\tilde{T}$-averaged $S_z^{rot}(t)$, their detuning robustness $\Delta_0$ is naturally limited by the same amount. As a sequence which avoids this limitation and reaches robustness to both dipolar fields and detuning that is not limited by the system-internal $J$ scale we introduce MREV-PulsePol.

\subsection*{LG-SLIC} \label{sec:seq_LGSLIC}
    Lee-Goldburg SLIC relies on a strong continuous wave Lee-Goldburg drive with full Rabi amplitude $\Omega(t)=\Omega$ to suppress dipolar coupling and adds a modulating drive which is equivalent to SLIC but applied in the Lee-Goldburg co-rotating frame. In \cite{dagysRobustParahydrogeninducedPolarization2024a}, this sequence was developed and demonstrated in an adiabatic version for homonuclear PHIP. For our purposes, the continuous-wave pulse can be described as
    \begin{align*}
        H_{ctrl}(t) &= (\vec{\Omega}_{LG}(t)+\vec{\Omega}_{mod}(t))\cdot\vec{S} \\
        \vec{\Omega}_{LG}(t)\cdot\vec{S} &= \Omega\ S_x + \Omega/\sqrt{2}\ S_z \\ 
            & =: \omega_{LG} \begin{pmatrix}
                \sqrt{2/3} \\  0 \\ \sqrt{1/3}
                \end{pmatrix} \cdot \vec{S}, \\
        \vec{\Omega}_{mod}(t)\cdot\vec{S} &= 2 J \cos(\omega_{LG}t)S_y,
    \end{align*}
    where $\vec{\Omega}_{mod}(t)$ fulfils the SLIC-condition by precessing the average $S_z^{rot}(t)$ created by the Lee-Goldburg drive. The latter uses a detuning $\Delta(t)=\Omega/\sqrt{2}$ and overall Rabi frequency $|\vec{\Omega}_{LG}|=\omega_{LG}$. As the polarisation relevant precession is oriented around the $y$-axis, we add suitable $\pi/2$ pulses to ensure overall $z$-alignment:

    \begin{align*}
    \pulse{\pi/2}{X}\left[  \pulse{CW}{X,\Delta(t)}^{\Omega(t)} \right]^N  \pulse{\pi/2}{\bar{X}}.
    \end{align*}
    Here, our definition slightly overshoots the maximum available Rabi frequency as $\vec{\Omega}_{mod}$ is added to $\vec{\Omega}_{LG}$ which we accept for the sake of a simpler definition. Similarly for the phase of the CW pulse which is only approximately given by $X$.
    As duration for one "repetition" we choose a full LG-rotation of $t_{\text{pulse}}=2\pi/\omega_{LG}$. 

\begin{figure}
		\centering
		\includegraphics[width=0.7\linewidth]{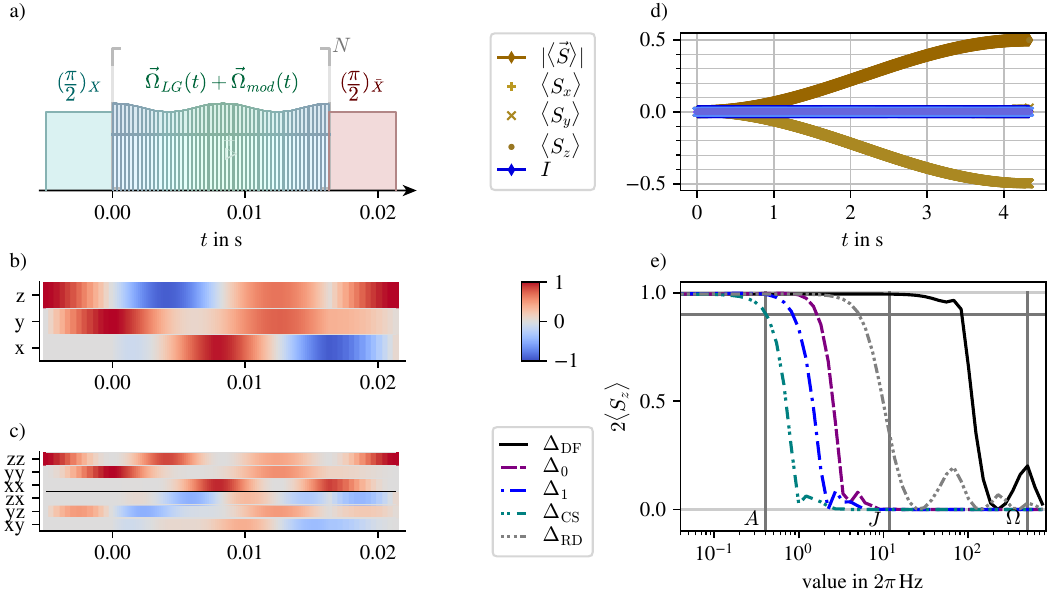}
		\caption{Characterisation of the LG-SLIC sequence as described in \Cref{sec:seq_describing_figs}. 
        a--c) use $\Omega=(2\pi)\,50\,$Hz for better readability.}
		\label{fig:seq_LGSLIC}
	\end{figure}

    In \Cref{fig:seq_LGSLIC}, we see that LG-SLIC indeed provides a strong dipolar field robustness, although the robustness to other errors is limited. Robustness to these errors may be increased to some degree by generalising the sequence to 
    frequency switched \cite{BieleckiKL1989},
    phase-modulated Lee-Goldburg  \cite{vinogradovHighresolutionProtonSolidstate1999} or LG4 decoupling \cite{HalseEmsley} but as these sequences follow similar principles we do not discuss them explicitly.

\subsection*{BLEWpol} \label{sec:seq_BLEWpol}

\begin{figure}
		\centering
		\includegraphics[width=0.7\linewidth]{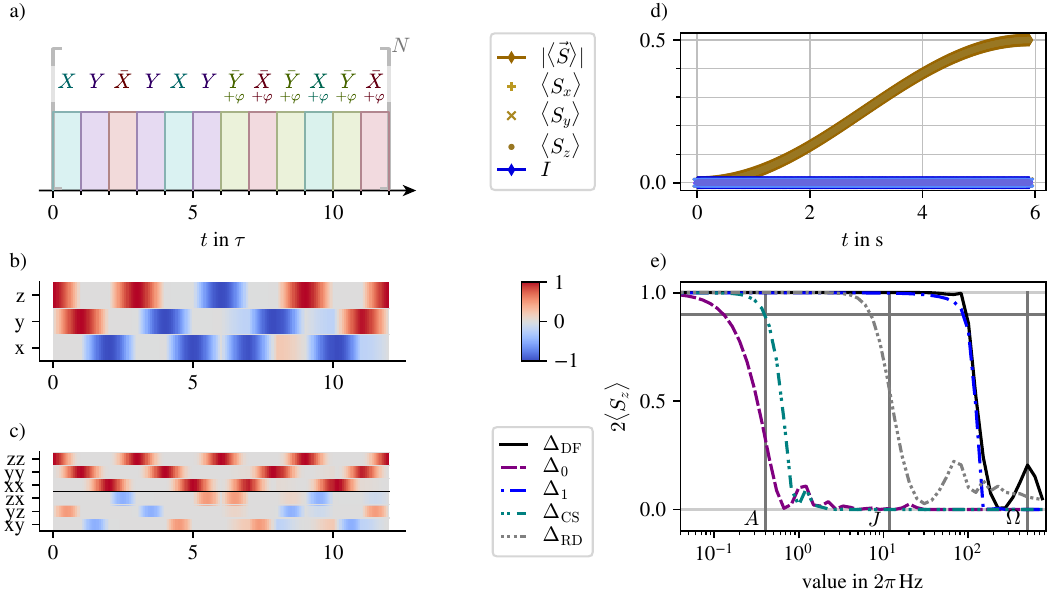}
		\caption{Characterisation of the BLEWpol sequence as described in \Cref{sec:seq_describing_figs}. 
        }
		\label{fig:seq_BLEWpol}
	\end{figure}

    Polarising BLEW-12 (BLEWpol) can be defined as 
    \begin{align*}
    \left[ \wait{\tau/2} \pulse{\pi/2}{X} \wait{\tau} \right. & \pulse{\pi/2}{Y} \wait{\tau} \pulse{\pi/2}{\bar{X}} \wait{\tau}  \pulse{\pi/2}{Y} \wait{\tau} \pulse{\pi/2}{X} \wait{\tau} \pulse{\pi/2}{Y} \wait{\tau}  \\
    \pulse{\pi/2}{\bar{Y}+\varphi} &\wait{\tau}  \pulse{\pi/2}{\bar{X}+\varphi} \wait{\tau}  \pulse{\pi/2}{\bar{Y}+\varphi} \wait{\tau} 
    \pulse{\pi/2}{X+\varphi} \wait{\tau} \left. \pulse{\pi/2}{\bar{Y}+\varphi} \wait{\tau}  \pulse{\pi/2}{\bar{X}+\varphi} \wait{\tau/2}\right]
    \end{align*}
    with the resonance condition $12\tau =\alpha = 2\varphi/J$ where the non-polarising BLEW-12 is given by $\varphi=0$. We can define the filling factor as $\eta=\dfrac{\pi}{2\Omega}\dfrac{1}{\tau}$. We choose the "windowless" $\eta=1$ where $S_z^{rot}(t)$ averaged over a single sequence repetition for $\varphi=0$ becomes \cite{burumLowpowerMultipulseLine1981} $2\sqrt{5}/(3\pi)(\sqrt{4/5}S_x+\sqrt{1/5}S_z)$. The non-vanishing $z$-component means that adding a $z$-precession via $\varphi\neq 0$ is not an ideal choice for creating a polarisation sequence. First, this wastes some of the potentially available effective coupling, and second, the overall net $z$-contribution creates a strong susceptibility to $\Delta_0$ errors which are not suppressed even in first order AHT of the $J$-precession time scale. BLEWpol has $A_\ast=4/(3\pi)\ A \approx 0.42\ A$. Correspondingly, the robustness evaluations in \Cref{fig:seq_BLEWpol} and \Cref{tab:seq_overview} show strong $\Delta_{DF}$ and $\Delta_1$ handling, but very high susceptibility to $\Delta_0$ errors. Panel c) shows the time-symmetry in the quadratic $S_z^{rot}(t)$ contributions which ensures second-order suppression of the dipolar coupling terms on the time scale of $\tilde{T}=12\tau=6\pi/\Omega$.
    As an example for a short pulsed dynamical decoupling sequence that can be easily modified to become polarising without as strong a $\Delta_0$ susceptibility, we present MREVpol in the next section.

\subsection*{MREVpol} \label{sec:seq_MREVpol}
    Polarising MREV-8 (MREVpol) is based on the non-polarising MREV-8 from \cite{rhimEnhancedResolutionSolid1973, mansfieldSymmetrizedMultipulseNuclearMagneticResonance1973} 
    which gives a good suppression of dipolar fields, the effect of finite pulse durations as well as corrections to $\Delta_1$ errors \cite{rhimAnalysisMultiplePulse1974}. Averaged over one repetition, $S_z^{rot}(t)$ lies on the $x$-$y$-plane such that we can create a polarisation sequence by adding a $z$-precession. With this understanding, we define Polarising MREV-8 as
    
    \begin{align*}
    \left[ \pulse{\pi/2}{X} \wait[1.5em]{2\tau} \pulse{\pi/2}{X} \wait{\tau} \right.&\left.\pulse{\pi/2}{Y} \wait[1.5em]{2\tau}\pulse{\pi/2}{\bar{Y}} \wait{1\tau}\right. \\
     \pulse{\pi/2}{\bar{X}+\varphi} &\wait[1.5em]{2\tau} \left.  \pulse{\pi/2}{\bar{X}+\varphi} \wait{\tau} \pulse{\pi/2}{Y} \wait[1.5em]{2\tau}\pulse{\pi/2}{\bar{Y}} \wait{1\tau} \right]^N
    \end{align*}
    with the resonance condition $12J \tau = \alpha = 2\varphi$ for the added precession to match the $I$-precession given by $J$ in Eq. \eqref{eq:H_rot}. For $\varphi=0$ we recover the original non-polarising MREV-8. It is helpful to define the filling factor $\eta = \dfrac{4\pi}{12\Omega\tau} = \dfrac{2\pi J}{\Omega\varphi}$ which can be freely chosen with $\eta \le 2/3$. Increasing $\eta$ strengthens the dipolar field suppression as the time scale of suppression decreases, and in return more repetitions $N$ are needed to complete the transfer. 
    Assuming $\Omega\gg J$ we have $\varphi \ll 1$ and in this limit $A_\ast = [\sqrt{2}/3+\eta(\frac{\sqrt{2}}{\pi}-\frac{\sqrt{2}}{4})]A\approx [0.47+0.1\,\eta]A$ with $A/A_\ast$ the chemical shift scaling factor of MREV-8 \cite{rhimEnhancedResolutionSolid1973}.

\begin{figure}
		\centering
		\includegraphics[width=1\linewidth]{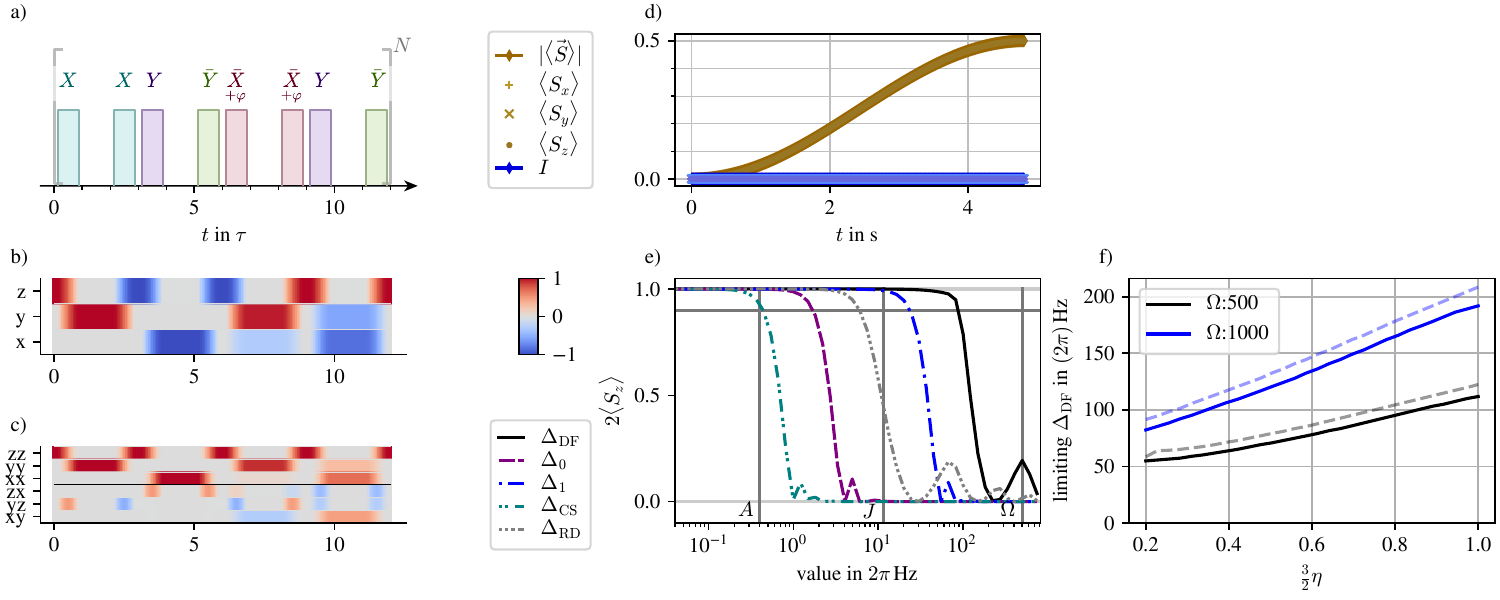}
		\caption{Characterisation of the MREVpol sequence with $3\eta/2=3/4$ as described in \Cref{sec:seq_describing_figs}. 
        Panel f) shows the limiting dipolar field $\Delta_{DF}$ for a variety of filling factors $\eta$ and different available Rabi amplitudes $\Omega$. The solid (dashed) lines correspond to the limit where $90\,$\% ($80\,$\%) of full transfer are reached.}
		\label{fig:seq_MREVpol}
	\end{figure}

    \Cref{fig:seq_MREVpol} shows the properties of MREVpol for $3\eta/2=3/4$ which are a strong suppression of dipolar fields $\Delta_{DF}$ combined with good robustness to amplitude errors $\Delta_1$. The robustness to detuning $\Delta_0$ fits the AHT estimate from precessing, and thus averaging to zero, the $\tilde{T}$-averaged $S_z^{rot}(t)$ on the time scale $T=2\pi/J$. 
    Next, we will introduce MREV-PulsePol which integrates MREV into PulsePol to reach simultaneous robustness to both $\Delta_{DF}$ and $\Delta_{0}$ that is only limited by the strength of the available control fields.

\subsection*{MREV-PulsePol} \label{sec:seq_MREVPP}
\begin{figure}
		\centering
		\includegraphics[width=1\linewidth]{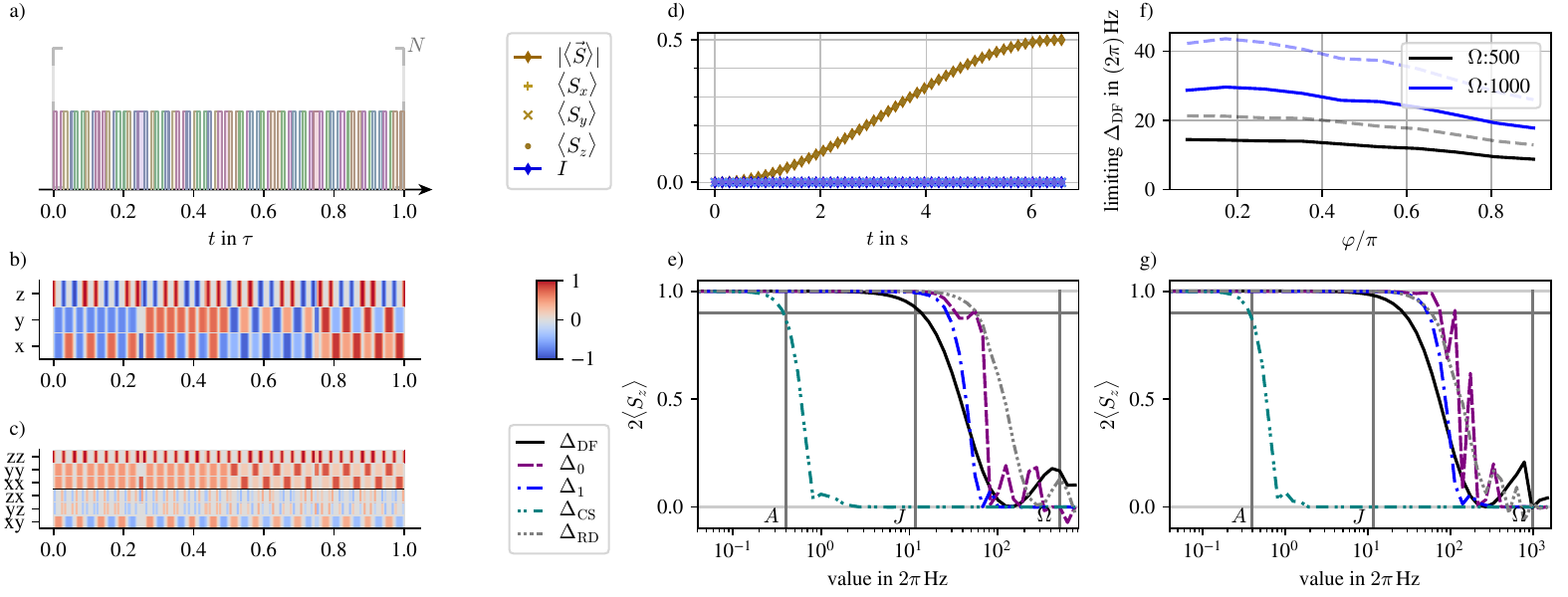}
		\caption{Characterisation of the MREV PulsePol sequence with $\varphi=0.4\pi$ and $J\tau=3.2\pi$ as described in \Cref{sec:seq_describing_figs}. 
        Panels a-c) use $\Omega=(2\pi)\,200\,$Hz and $m=2$ for readability.
        Panel f) shows the limiting dipolar field $\Delta_{DF}$ for a variety of phases $\varphi$ and different available Rabi amplitudes $\Omega$. The solid (dashed) lines correspond to the limit where $90\,$\% ($80\,$\%) of full transfer are reached.
        Panel g) shows the single error robustness as in e) but for an increased Rabi amplitude of $\Omega=(2\pi)\, 1000\,$Hz.
        }
		\label{fig:seq_MREVPP}
	\end{figure}

    While sequences like LG-SLIC and MREVpol provide very strong suppression of dipolar fields, their robustness especially to detuning errors $\Delta_0$ is limited by the scale of $J$. To overcome this limitation, MREV-PulsePol uses MREV on the quick $\tilde{T}$ time scale to suppress the dipolar interaction and, instead of directly following the $J$-precession with $S_z^{rot}(t)$ on the slower $T\sim 1/J$ time scale, uses the corresponding $S_z^{rot}(t)$ trajectory of PulsePol.
    We use the shorthand notation $\underset{\text{MREV}^m_\phi}{\wait[3em]{\tau_\text{free}}}$ for $m$ repetitions of the MREV sequence with $\varphi=0$ and a total duration of $\tau_{\text{free}}$ and a global shift in all pulse phases by $\phi$ such that 
     \begin{align*}
    \underset{\text{MREV}^m_\phi}{\wait[3em]{\tau_\text{free}}} := & \left[ \pulse{\pi/2}{X+\phi} \wait[2.5em]{\tau_\text{free}/(6m)} \pulse{\pi/2}{X+\phi} \wait[2em]{\tau_\text{free}/(12m)}  \pulse{\pi/2}{Y+\phi} \wait[2.5em]{\tau_\text{free}/(6m)}\pulse{\pi/2}{\bar{Y}+\phi} \wait[2em]{\tau_\text{free}/(12m)} \right. \\
     & \pulse{\pi/2}{\bar{X}+\phi} \wait[2.5em]{\tau_\text{free}/(6m)} \left.  \pulse{\pi/2}{\bar{X}+\phi} \wait[2em]{\tau_\text{free}/(12m)} \pulse{\pi/2}{Y+\phi} \wait[2.5em]{\tau_\text{free}/(6m)}\pulse{\pi/2}{\bar{Y}+\phi} \wait[2em]{\tau_\text{free}/(12m)} \right]^m.
    \end{align*}
    When starting with $S_z^{rot}(t)=+S_z$, $U^{rot}(t)=\mathbb{1}$ and applying $\underset{\text{MREV}^m_{3\pi/4}}{\wait[3em]{\tau_\text{free}}}$, the average orientation over that period is $\frac{A}{\tau_\text{free}}\int_0^{\tau_{\text{free}}}\mathrm{d}t'\ S_z^{rot}(t+t')=A_\ast^{\text{MREV}}S_x$ with the final orientation returning to $S_z^{rot}(t+\tau_{\text{free}})=+S_z$, $U^{rot}(t+\tau_\text{free})=U^{rot}(t)$. With this as a dipolar decoupling building block, we use $\underset{\overline{\text{MREV}}^{m}_\phi}{\wait[3em]{\tau_\text{free}}}$ to denote the above in time-reverse order and define MREV-PulsePol as 
    \begin{align*}
        \left[ \underset{\text{MREV}^m_{3\pi/4}}{\wait[3em]{\tau/4}} \pulse{\pi}{Y} \underset{\overline{\text{MREV}}^{m}_{3\pi/4}}{\wait[3em]{\tau/4}} 
            \underset{\text{MREV}^m_{3\pi/4+\varphi}}{\wait[3em]{\tau/4}} \pulse{\pi}{Y+\varphi} \underset{\overline{\text{MREV}}^{m}_{3\pi/4+\varphi}}{\wait[3em]{\tau/4}} \right]^N.
    \end{align*}
    Note that the averaged $S_z^{rot}(t)$ orientation during the MREV parts ensures that no additional $\pi/2$ pulses are required to reach the $J$ time scale behaviour of PulsePol.
    Due to this, the resonance condition remains identical to that of PulsePol with the strongest resonance for $J\tau = 4\pi -2\varphi$.
    For a strong dipolar field suppression, we choose the largest number of $m$ that still allows for a filling factor $\eta$ below $90\%$ of the maximum to ensure enough time for the $\pi$ pulses, and in the limit of large $m$ the effective coupling is given by $A_\ast = A_\ast^{\text{PulsePol}}A_\ast^{\text{MREV}}/A$. Each sequence repetition creates a net precession of $\alpha=2\varphi$.
    For \Cref{fig:seq_MREVPP} we choose $\varphi=0.4\pi$ with $m=5$ and $A_\ast \approx 0.38\ A$, where panels b,c) on short time scales show the structure of MREV from \Cref{fig:seq_MREVpol} and on longer time scales the structure of PulsePol from \Cref{fig:seq_PulsePol}.
    Except for the chemical shift error $\Delta_{CS}$ which remains unaffected by any pulses on the $S$ spin, and $\Delta_{RD}$ for which the robustness is very strong, but still fundamentally tied to $J$, MREV-PulsePol exceeds the $J$-scale for all errors in terms of robustness. The robustness to pulse errors $\Delta_{0/1}$ is only slightly weaker than for DF-PulsePol and dipolar field suppression at the reference Rabi amplitude (panel e)) is significantly weakened compared to e.g. MREVpol while still surpassing the other PulsePol based sequences. Panel g) shows the single error robustness for $J\to J/4$ and an otherwise identical setting, and clearly demonstrates that the robustness of MREV-PulsePol to these errors is not limited by the $J$-scale. 

	\clearpage
	\section{Dual channel control}\label{sec:dual_channel}

    In the previous sections we have examined dipolar field decoupling sequences with varying levels of robustness to different imperfections. However, achieving this robustness is necessarily accompanied with a reduction of the transfer rate $A_\ast$ by a factor of at least $\sqrt{2/3}$ to fulfil \cref{eq:DF_suppressing}. Similarly, the refocusing structure of PulsePol, which provides strong detuning robustness $\Delta_0$, leads to an effective coupling strength reduced by a factor $\le 0.73$. Hence, while highly robust sequences are possible, they are inevitably associated with a significant reduction in the transfer rate. This trade-off may be acceptable when the relevant polarisation lifetimes are long, but for shorter lifetimes, it directly limits the achievable polarisation levels.
    Moreover, these sequences offer limited robustness to differential chemical shifts $\Delta_{CS}$ which scale with the applied magnetic field $B_0$. This limits the maximally usable $B_0$, particularly for molecules where relevant hydrogen nuclei are not fully chemically equivalent. As none of the sequences discussed so far involve active control of the hydrogen nuclei, they are all subject to this limitation (cf.~\Cref{tab:seq_overview}).
    
    In this section, we introduce sequences which apply pulses to both of the spin types which allows for ameliorating 
    both of these limitations. We first describe the changes that regular $\pi$ pulses on the hydrogen spins create for the dynamics and conditions on polarisation sequences. With that, we introduce PulsePol+XY as a faster version of PulsePol which, however, remains susceptible to dipolar fields. We then continue with the sequence generalising M2A-PulsePol, which gains slightly in its suppression of dipolar fields and DF-PulsePol+XY which pushes towards the limits of $J$-scale suppression. Finally, we present alternating MREVpol+XY
    which is a suitable adaptation of MREVpol which is formally limited in its suppression of all errors only by the available Rabi amplitudes while retaining a higher transfer rate $A_\ast$ than MREV-PulsePol. However, despite this formal advantage, at any fixed Rabi amplitude, the more dedicated sequences reach significantly stronger robustness for those errors that they are designed to handle well. 
    Nevertheless, there remains the possibility that sequences can be found that show much less of a gap even while incorporating correction to all regarded errors.
    
    \subsection*{The effect of I spin \texorpdfstring{$\pi$}{pi} pulses} \label{sec:2ch_seqs_theory}
    In this section we initially assume ideal pulses with vanishing duration to retain a description for polarisation sequences that relies only on the $S$ spin and the $I$ pseudo-spin. The numerical results do not rely on this approximation.

    For this, we first note that a $\pi$ pulse $\pulse{\pi}{X}$ using $\vec{\Omega}_{I}(t)$ corresponds to the unitary $U_{\pi}^I=\exp(-i\pi (I_{1,x}+I_{2,x}))$, which transforms the $I_i$ containing terms of our Hamiltonian Eq.~\eqref{eq:H} as
    \begin{align}\label{eq:2ch_pi_effect}
        U_{\pi}^I \left(\vec{I}_1\cdot \vec{I}_2\right) U_{\pi}^{I\dagger} &=  \vec{I}_1\cdot \vec{I}_2 \\ \nonumber
        U_{\pi}^I  \left(I_{1,z} \pm I_{2,z}\right) U_{\pi}^{I\dagger} &= -1\cdot \left(I_{1,z} \pm I_{2,z}\right) \\ \nonumber
        \Rightarrow H(J,A,A^\Sigma,\Delta_{CS}) &\overset{U_\pi^I}{\rightarrow} H(J,-A,-A^\Sigma,-\Delta_{CS}).
    \end{align}
    Due to $z$-rotational symmetry, this effect does not depend on the phase of the $\pi$-pulse.
    Thus, $\pi$ pulses on the $I_{i}$ spins effectively flip the sign of our transfer coupling $A$, as well as the sign of $A^\Sigma$ and $\Delta_{CS}$.
    Assuming $\Omega_{I}(t)$ to correspond to a CPMG-like sequence with a $\pi$ pulse every $T^I$, we can introduce the filter function $f^I(t)=\sgn(\sin(\pi\ t/T^I))$
    to reach an adjusted Eq. \eqref{eq:H_rot} with
    \begin{align} \label{eq:H_rot_2ch}
        H^{rot}(t) = A f^I(t) S_z^{rot}(t) (I_x\cos(Jt)-I_y \sin(Jt))+ H_{err}^{rot}(t),
    \end{align}
    where $H_{err}^{rot}(t)$ is defined as in the previous sections except for $\Delta_{CS}\rightarrow f^I(t)\Delta_{CS}$ which naturally averages to zero in the effective Hamiltonian. Thus, 2-channel sequences naturally suppress differential chemical shift on the time scale $T^I$.
    Polarisation sequences now need to take into account that $f^I(t)$ effectively flips $S_z^{rot}(t)$ every $T^I$ compared to the previous sections. We will see that this can be advantageous for sequence properties.

    The above description relies on the assumption of negligibly short pulses. However, in order to suppress dipolar fields on a quick time scale $\tilde{T}$, it is specifically the regime with short $T^I$ which is of interest. Here, a significant fraction of the overall duration is filled with pulses such that using $f^I(t)$ is no longer sufficient to describe all of the properties. Numerically, we have found that specifically the presence of dipolar fields becomes more challenging to sequences in this regime where the dynamics during active $I$-pulses can create hydrogen magnetisation which is associated with significantly stronger dipolar fields than for carbon spins. The numerically evaluated robustness properties of sequences presented in this section include these effects, however a detailed theoretical exploration exceeds the scope of this work.

    
    \subsection*{PulsePol+XY} \label{sec:seq_PP_XY}
    For PulsePol+XY, we combine the PulsePol sequence from \Cref{sec:seq_PulsePol} with a synchronised CPMG sequence, with XY4-like phases but parametrised by $\varphi_I$:
    
    \begin{align*}
    \vec{\Omega}:\qquad &\left[ \pulse{\pi/2}{X} \wait{\tau/4} \pulse{\pi}{Y} \wait{\tau/4} \pulse{\pi/2}{X} \cdot \pulse{\pi/2}{X+\varphi} \wait{\tau/4} \pulse{\pi}{Y+\varphi} \wait{\tau/4} \pulse{\pi/2}{X+\varphi} \right]^N \\
    \vec{\Omega}_{I}:\ \pulse{\pi/2}{X} & \left[ \pulse{\pi/2}{X} \wait{\tau/4} \pulse{\pi}{\varphi_I} \wait{\tau/4} \pulse{\pi/2}{X} \cdot \pulse{\pi/2}{X} \wait{\tau/4} \pulse{\pi}{\varphi_I} \wait{\tau/4} \pulse{\pi/2}{X} \right]^N  \pulse{\pi/2}{X}
    \end{align*}
    
    The resonance condition is unaffected except for a flip $\varphi\to-\varphi$ such that $\pm\hat{e}_z$ magnetisation is created at $J\tau=2n(2\pi)\mp 2\varphi$, but the transfer speed is now $A_\ast=A \dfrac{\sin(J\tau/4)}{J\tau/4}$ which leads to the natural choice for the resonance becoming $n=0$. This allows for quicker repetition rates $J\tau\ll 1$ with $A_\ast$ approaching $A$. Each sequence repetition creates a net precession of $\alpha=2\varphi$. We define the filling factor $\eta= \Omega\tau/(4\pi)=\varphi/(2\pi) \times\Omega/J$

\begin{figure}
		\centering
		\includegraphics[width=\linewidth]{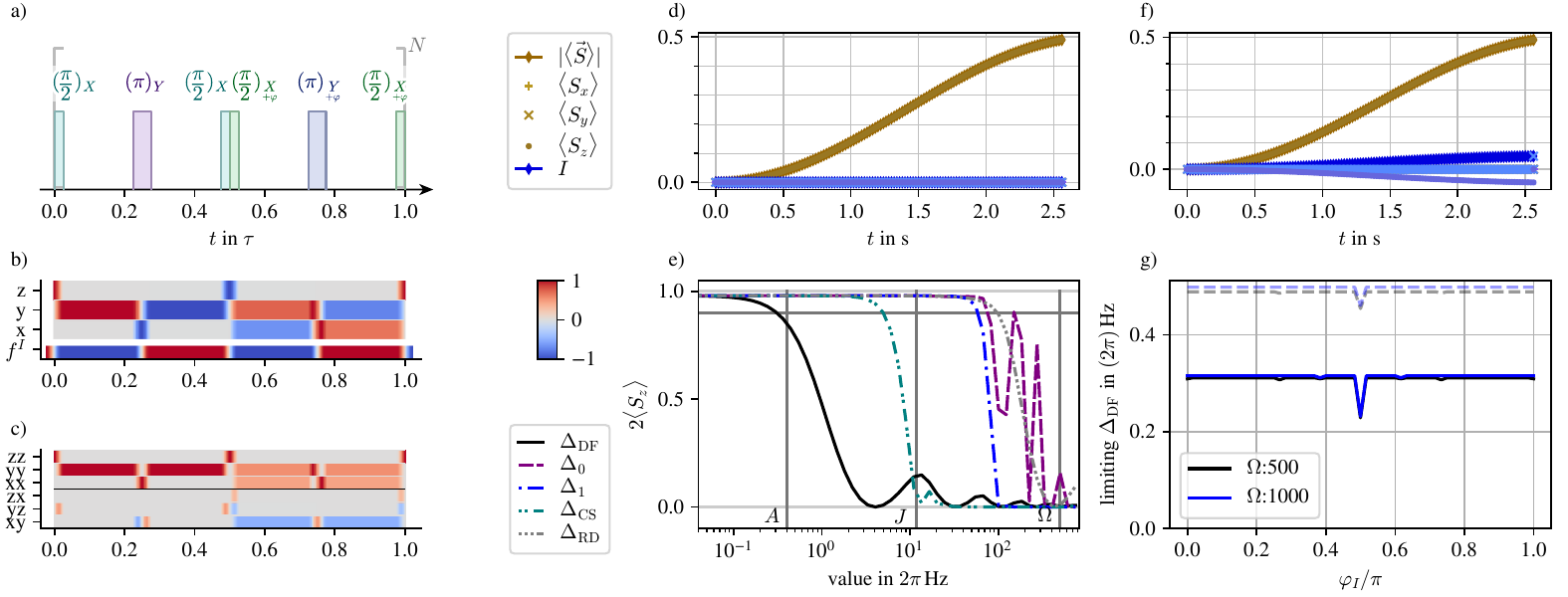}
        \caption{Characterisation of the PulsePol+XY sequence with $\eta=1/5$ and $\varphi_I=3\pi/4$ as described in \Cref{sec:seq_describing_figs}.
        Panel f) shows the time evolution for $\varphi_I=\pi/2$ and panel g) gives the limiting dipolar field strength for a variety of $\varphi_I$ choices.
        }
		\label{fig:seq_PP_XY}
	\end{figure}

    In \Cref{fig:seq_PP_XY} we see that indeed PulsePol+XY reaches higher $A_\ast$ and suppresses $\Delta_{CS}$ thanks to the hydrogen pulses. 
    We choose $\eta=1/5$ which is sufficiently high to demonstrate these effects, while being sufficiently low that the effective coupling $A_\ast$ given above remains an acceptable approximation, given that it assumes ideal pulses and thus $\eta \ll 1$. The parameter $\varphi_I$ is chosen as $3\pi/4$ as we find in panel g) that the standard XY4 choice of $\varphi_I$ leads to reduced $\Delta_{DF}$ robustness. For this value, panel f) shows that there is a small, but significant build-up of $I$ spin magnetisation which due to the higher gyromagnetic ratio is associated with a significantly stronger dipolar field.
    For the chosen $\varphi_I$, we see no such build-up (cf. panel d)) and while dipolar fields ($\Delta_{DF}$) are not suppressed, the increased transfer speed by itself leads to a small improvement compared to PulsePol. 
    The importance of $\varphi_I$ suggests that corrections to our assumption of instantaneous pulses in deriving \Cref{eq:H_rot_2ch} play an important role in the numerically evaluated robustness properties.
    
    \clearpage
    
	\subsection*{M2A-PP+XY} \label{sec:seq_M2A_PP_XY}
    The approach of PulsePol+XY can be similarly used for all of the dipolar field variations of PulsePol which we abbreviate as "PP" for shorter sequence names.  
    In particular, the version with M2A-PulsePol takes the form
    \begin{align*}\label{eq:seq:M2A_PulsePol_XY}
    \vec{\Omega}:\quad \left[ \pulse{\pi/4}{X} \right. \wait[2em]{\tau/4} \pulse{\pi}{Y} \wait[2em]{\tau/4} \pulse{\pi/4}{X} &\cdot \pulse{\pi/2}{X+\varphi} \wait{\tau/4} \pulse{\pi}{Y+\varphi} \wait{\tau/4} \pulse{\pi/2}{X+\varphi} \\ \nonumber 
     \pulse{\pi/4}{X} \wait[2em]{\tau/4} \pulse{\pi}{Y} \wait[2em]{\tau/4} \pulse{\pi/4}{X} &\cdot \pulse{\pi/4}{X+\varphi} \wait[0.5em]{\tau/4} \pulse{\pi}{Y+\varphi} \wait[0.5em]{\tau/4} \pulse{\pi/4}{X+\varphi} \\ \nonumber 
     \pulse{\pi/2}{X} \wait{\tau/4} \pulse{\pi}{Y} \wait{\tau/4} \pulse{\pi/2}{X} &\cdot \pulse{\pi/4}{X+\varphi} \wait[2em]{\tau/4} \pulse{\pi}{Y+\varphi} \wait[2em]{\tau/4} \left. \pulse{\pi/4}{X+\varphi}  \right]^N \\ \nonumber
     \vec{\Omega}_{I}:\quad   \pulse{\pi/2}{X}\left[ \pulse{\pi/2}{X} \wait{\tau/4} \pulse{\pi}{\varphi_I} \wait{\tau/4} \pulse{\pi/2}{X} \right. & \left. \cdot \pulse{\pi/2}{X} \wait{\tau/4} \pulse{\pi}{\varphi_I} \wait{\tau/4} \pulse{\pi/2}{X} \right]^{3N}  \pulse{\pi/2}{X}
    \end{align*}

    where the resonance condition remains that of PulsePol+XY, and the effective transfer speed assuming ideal pulses is given by $A_\ast=A \ \dfrac{1+2\sqrt{1/2}}{3} \dfrac{\sin(J\tau/4)}{J\tau/4}$. The adjacent waiting times correspondingly are adjusted by their respective "$\pi/2$" pulse duration. Each sequence repetition creates a net precession of $\alpha=6\varphi$.
    
\begin{figure}
		\centering
		\includegraphics[width=\linewidth]{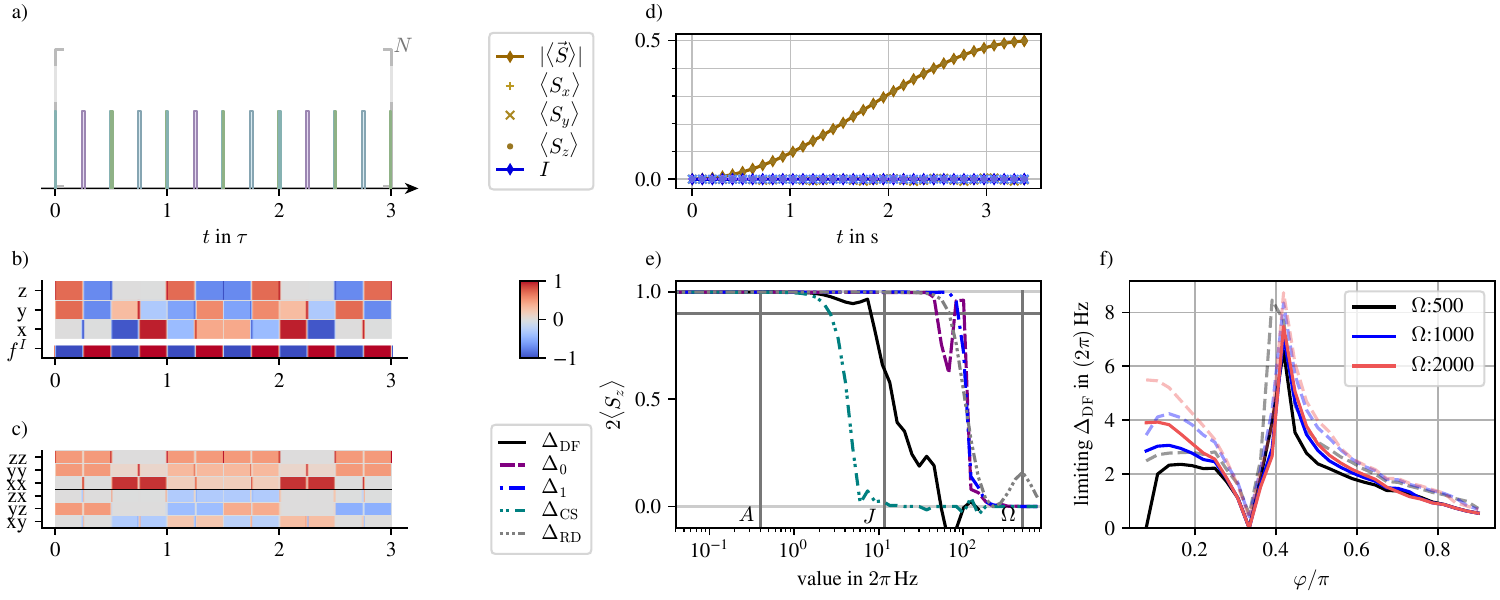}
        \caption{Characterisation of the M2A-PP+XY sequence with $\varphi=0.4\pi$, $\varphi_I=3\pi/4$ and $J\tau=2\varphi$ as described in \Cref{sec:seq_describing_figs}. 
        f) shows the limiting dipolar field $\Delta_{DF}$ for a variety of phases $\varphi$ using $J\tau=2\varphi$ and different available Rabi amplitudes $\Omega$. The solid (dashed) lines correspond to the limit where $90\,$\% ($80\,$\%) of full transfer are reached.
        }
		\label{fig:seq_M2APP_XY}
	\end{figure}

    Illustration \Cref{fig:seq_M2APP_XY} e) shows the corresponding robustness to single errors with the choice $\varphi=0.4\pi=J\tau/2$ and $\varphi_I=3\pi/4$. Transfer in the presence of dipolar fields is essentially unaffected up to $\Delta_{DF}\sim \sqrt{A_\ast J}$ and strong transfer is retained up to values close to $J$. This is also the time scale of \Cref{eq:DF_suppressing} where the dipolar field terms are averaged to zero. Additionally, we see that the choice of $\varphi\approx 0.4\pi$ is important for good dipolar field suppression in panel f). 
    
	\subsection*{DF-PP+XY} \label{sec:seq_DF_PP_XY}
    For the dual channel version of DF-PulsePol, we find that dipolar field robustness is improved if the final part of the PulsePol repetitions uses a full $\pi$ rotation to replace the $\pi/2$ as compared to removing them as is preferable in DF-PulsePol as a single channel sequence. With this, DF-PP+XY is defined as

    \begin{align*}
    \vec{\Omega}:\quad     \left[ \pulse{\pi/2}{X} \right. \wait[1.4em]{\tau/4} \pulse{\pi}{Y} \wait[1.4em]{\tau/4} \pulse{\pi/2}{X} & \pulse{\pi/2}{X+\varphi} \wait[1.4em]{\tau/4} \pulse{\pi}{Y+\varphi} \wait[1.4em]{\tau/4} \pulse{\pi/2}{X+\varphi} \\ \nonumber 
     \wait[4.4em]{\tau/4} \pulse{\pi}{Y} \wait[4.4em]{\tau/4}  & \pulse{\pi/2}{X+\varphi} \wait[1.4em]{\tau/4} \pulse{\pi}{Y+\varphi} \wait[1.4em]{\tau/4} \pulse{\pi/2}{X+\varphi} \\ \nonumber
    \pulse{\pi/2}{X}\wait[1.4em]{\tau/4} \pulse{\pi}{Y} \wait[1.4em]{\tau/4} \pulse{\pi/2}{X}&\pulse{\pi}{X+\varphi} \wait{\tau/4} \left. \pulse{\pi}{Y+\varphi} \wait{\tau/4} \pulse{\pi}{X+\varphi} \right]^N\, \\
     \vec{\Omega}_{I}:\quad   \pulse{\pi/2}{X}\left[ \pulse{\pi/2}{X} \wait{\tau/4} \pulse{\pi}{\varphi_I} \wait{\tau/4} \pulse{\pi/2}{X} \right. & \left. \cdot \pulse{\pi/2}{X} \wait{\tau/4} \pulse{\pi}{\varphi_I} \wait{\tau/4} \pulse{\pi/2}{X} \right]^{3N}  \pulse{\pi/2}{X}.
    \end{align*}

    For DF-PP+XY, too, the resonance condition remains that of PulsePol+XY, and the effective transfer speed is now given by $A_\ast=A \ \dfrac{2}{3} \dfrac{\sin(J\tau/4)}{J\tau/4}$. The waiting times are each adjusted by their respective "$\pi/2$" pulse duration and each sequence repetition creates a net precession of $\alpha=6\varphi$.
    Illustration \Cref{fig:seq_M2APP_XY} e) shows the corresponding robustness to single errors for $\varphi=0.4\ \pi$ and $\varphi_I=3\pi/4$. For the chosen parameters, the $\Delta_{DF}$ strengths with near-optimal transfer reach $J$. In panel f) we see, that for small $\varphi$, DF-PP+XY can profit further from increased Rabi amplitudes which allow for strong transfer at dipolar fields comparable to or even exceeding $2J$. As with DF-PulsePol, the actual cancellation of dipolar field terms is limited to the $J$ time scale even if transfer can remain successful for somewhat stronger dipolar fields.
    
	\begin{figure}
		\centering
		\includegraphics[width=\linewidth]{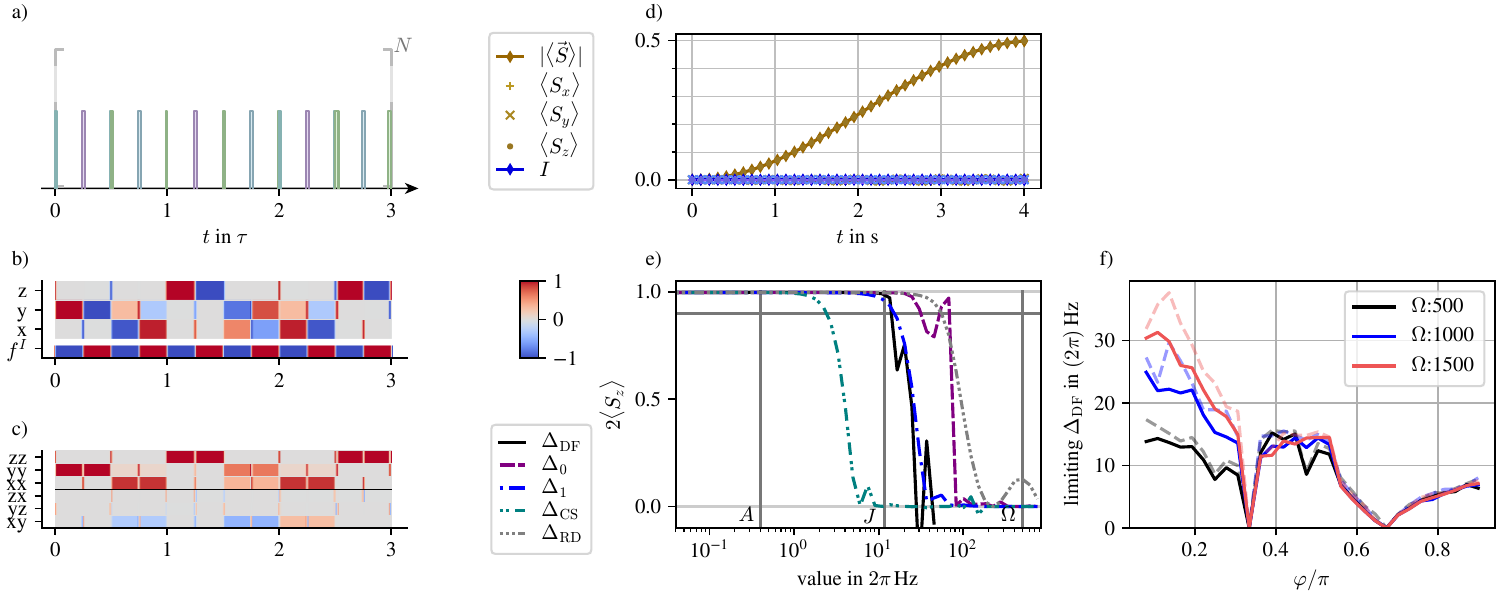}
        \caption{Characterisation of the DF-PP+XY sequence with $\varphi=0.4\pi$, $\varphi_I=3\pi/4$ and $J\tau=2\varphi$ as described in \Cref{sec:seq_describing_figs}. 
        f) shows the limiting dipolar field $\Delta_{DF}$ for a variety of phases $\varphi$ using $J\tau=2\varphi$ and different available Rabi amplitudes $\Omega$. The solid (dashed) lines correspond to the limit where $90\,$\% ($80\,$\%) of full transfer are reached.
        }
		\label{fig:seq_DF_PP_XY}
	\end{figure}

	\subsection*{altMREVpol+XY} \label{sec:seq_altMREVpol_XY}

\begin{figure}
		\centering
		\includegraphics[width=\linewidth]{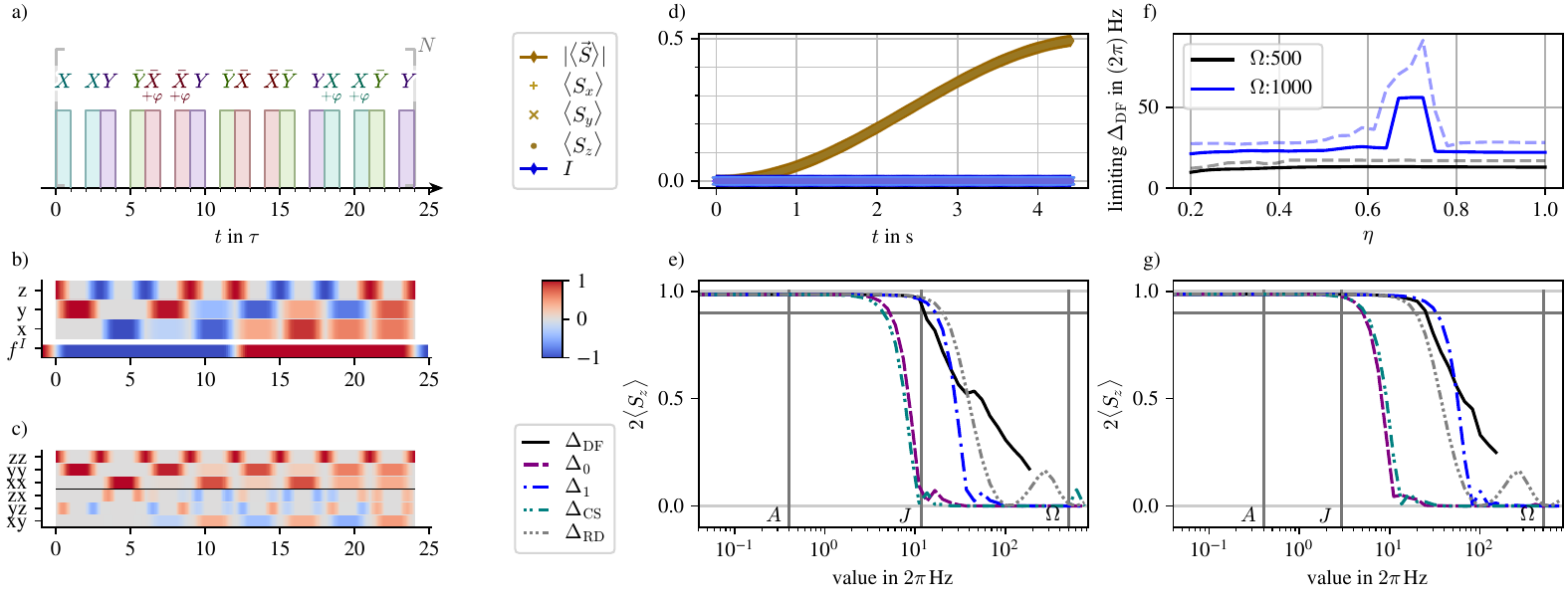}
        \caption{Characterisation of the altMREVpol+XY sequence with $\eta=2/3$, $\varphi_I=0.9\pi$ as described in \Cref{sec:seq_describing_figs}. 
        f) shows the limiting dipolar field $\Delta_{DF}$ for a variety of phases $\varphi$ using $J\tau=2\varphi$ and different available Rabi amplitudes $\Omega$. The solid (dashed) lines correspond to the limit where $90\,$\% ($80\,$\%) of full transfer are reached.
        g) shows the single parameter robustness as in e) but uses $J\to J/4$.}
		\label{fig:seq_altMREVpol_XY}
	\end{figure}

    Alternating MREVpol improves on the $\Delta_0$ robustness of MREVpol by using two cycles of MREVpol which alternate between opposing phases. A $\pi$ pulse on the $I$ spins ensures that the alternating phases on the $S$ spin do not further affect the effective interaction:
    \begin{align*}
    \vec{\Omega}:\quad &    \left[ \pulse{\pi/2}{X} \,\;\wait[1.5em]{2\tau}\;\, \pulse{\pi/2}{X} \,\;\wait{\tau}\;\, \pulse{\pi/2}{Y} \wait[1.5em]{2\tau}\pulse{\pi/2}{\bar{Y}} \wait{1\tau}\right. \\
    & \pulse{\pi/2}{\bar{X}+\varphi} \wait[1.5em]{2\tau}  \pulse{\pi/2}{\bar{X}+\varphi} \wait{\tau} \pulse{\pi/2}{Y} \wait[1.5em]{2\tau}\pulse{\pi/2}{\bar{Y}} \wait{1\tau} \\
    &  \pulse{\pi/2}{\bar{X}} \,\;\wait[1.5em]{2\tau}\,\; \pulse{\pi/2}{\bar{X}} \,\,\;\wait{\tau}\;\, \pulse{\pi/2}{\bar{Y}} \wait[1.5em]{2\tau}\pulse{\pi/2}{Y} \wait{1\tau} \\
    & \pulse{\pi/2}{X+\varphi} \wait[1.5em]{2\tau} \left.  \pulse{\pi/2}{X+\varphi} \wait{\tau} \pulse{\pi/2}{\bar{Y}} \wait[1.5em]{2\tau}\pulse{\pi/2}{Y} \wait{1\tau} \right]^N \\
    \vec{\Omega}_{I}:\quad &  \pulse{\pi/2}{X} \left[ \pulse{\pi/2}{X} \wait[3em]{12\tau} \pulse{\pi}{\varphi_I} \wait[3em]{12\tau} \pulse{\pi/2}{X} \right]^N  \pulse{\pi/2}{X}
    \end{align*}
    In the limit of small $\varphi$, altMREVpol+XY has the same effective coupling as MREVpol with $A_\ast \approx [0.47+0.1\,\eta]A$ and the resonance condition is given by $12\tau J = \alpha=2\varphi$.
    In \Cref{fig:seq_altMREVpol_XY} the sequence with maximum filling factor $\eta=\dfrac{4\pi}{12\tau \Omega}=2/3$ is used to maximise the rate at which the phases alternate. Further, we choose $\varphi_I=0.9\pi$. Comparing the robustness properties to MREVpol directly, the detuning robustness ($\Delta_0$) is indeed strengthened together with the robustness to chemical shifts ($\Delta_{CS}$). The robustness to dipolar fields however is significantly reduced at the same Rabi amplitude. Panel f) indicates that higher Rabi amplitudes correspondingly improve the dipolar decoupling and it appears some choices of $\eta$ at the increased Rabi amplitudes further improve the $\Delta_{DF}$ suppression. Potentially, there is an ideal choice of $\varphi$ that is reached by these values. Panel g) shows the single-parameter robustness for $J\to J/4$ to demonstrate that for altMREVpol+XY, the robustness to all regarded errors is not limited by $J$, but is only limited by $\Omega$.

\section{Conclusion}\label{sec:conclusion}

Dipolar fields can be a challenge for transfer in high-concentration PHIP samples. While typically polarisation sequences are susceptible to dipolar fields (\cref{sec:unadjusted}), there are alternatives which strongly improve the robustness to $\Delta_{DF}$ without significantly sacrificing other properties: with slightly lowered transfer speed, moderate $\Delta_{DF}$ can be overcome with comparable pulse-error robustness ($\Delta_{0}$ and $\Delta_{1}$, \cref{sec:DFsuppressing_(1)}). Strong $\Delta_{DF}$ beyond $J$ can be suppressed with specialized dipolar decoupling sequences which applied directly have $J$-limited detuning robustness, but allow for combination with the principle of PulsePol to overcome this limitation. Sequences which drive both spins pose requirements on B0 and selectivity, but in principle also allow for overcoming this robustness trade-off while additionally providing a speed benefit and allowing for the suppression of moderate chemical shifts. While our suggested altMREVpol+XY provides a proof-of-principle of strong suppression for both dipolar fields and all pulse errors, it needs very strong pulses to actually play out its benefits compared to the other sequences developed in this work. It seems plausible that a carefully designed dual channel sequence can be found which has the same strengths and becomes advantageous already at smaller B1 amplitudes.
Other directions of future research could be a more detailed understanding of the stabilising abilities of the dipolar field that were found in amplitude swept SLIC, carefully analysing MREV-PulsePol or a similar combination to retain a larger fraction of the dipolar field suppression while retaining the pulse error robustness, and a more detailed analysis of the effects of dipolar fields when both spins are driven and might accumulate magnetisation. 

\section{Acknowledgements} This work was supported by the ERC Synergy grant HyperQ (grant no. 856432), the BMBF projects QuE-MRT (grant no 13N16447) and QMED2-PHIP-NMR (grant no 03ZU2110CB) and the EU project SPINUS (grant no 101135699) and C-QuENS (grant no 101135359). The authors acknowledge fruitful discussions with Pol Alsina-Bolivar, Laurynas Dagys, Stephan Knecht, Vitaly Kozinenko and Bogdan Rodin.

\clearpage
\bibliography{DDpol_library.bib}

\clearpage
\appendix

\end{document}